\documentclass[10pt]{article}

\usepackage{amsmath}
\usepackage{amsfonts}
\usepackage{amssymb}
\usepackage[dvips]{graphicx}
\usepackage{subfigure}

\begin{document}

\newcommand{\be}{\begin{equation}}\newcommand{\ee}{\end{equation}}
\newcommand{\bea}{\begin{eqnarray}} \newcommand{\eea}{\end{eqnarray}}
\newcommand{\ba}[1]{\begin{array}{#1}} 
\newcommand{\ea}{\end{array}}
\newcommand{\beq}{\begin{equation}} \newcommand{\eeq}{\end{equation}}

\numberwithin{equation}{section}

\def\np{Nucl. Phys. {\bf B}}\def\pl{Phys. Lett. {\bf B}}
\def\mpl{Mod. Phys. {\bf A}}\def\ijmp{Int. J. Mod. Phys. {\bf A}}
\def\cmp{Comm. Math. Phys.}\def\prd{Phys. Rev. {\bf D}}

\def\a{\alpha}
\def\b{\beta}
\def\g{\gamma}
\def\w{\omega }
\def\A{\mathcal{A}}
\def\eps{\epsilon}
\def\p{\partial }
\def\vphi{\varphi}
\def\s{\sigma }
\def\nn{\nonumber }
\def\non{\nonumber }

\begin{flushright}
ICCUB-10-020\\
NSF-KITP-10-033 \\

March, 2010
\end{flushright}

\bigskip

\begin{center}

{\Large\bf  Phenomenological Models of Holographic Superconductors\\
 and  Hall currents }

\bigskip
\bigskip

{\it \large Francesco Aprile\,$^1$,  Sebasti\'an Franco\,$^2$, Diego Rodr\'iguez-G\'omez\,$^3$ and  Jorge G. Russo\,$^{1,4}$}
\bigskip

{\it
1) Institute of Cosmos Sciences and Estructura i Constituents de la Materia\\
Facultat de F{\'\i}sica, Universitat de Barcelona\\
Barcelona, Spain \\
\smallskip
2) KITP, University of California, Santa Barbara, CA93106-4030, USA \\
\smallskip
3) Queen Mary, University of London, Mile End Road, London E1 4NS, UK \\
\smallskip
4) Instituci\'o Catalana de Recerca i Estudis Avan\c cats (ICREA),\\
Pg. L. Companys, 23, 08010 Barcelona, Spain \\
}
\bigskip
\bigskip

\end{center}
\bigskip

\begin{center}

\begin{abstract}

We study general models of holographic superconductivity parametrized by four arbitrary functions of a neutral scalar field of the bulk theory.
The models can accommodate several features of real superconductors, like  
arbitrary critical temperatures and critical exponents in a certain range, and perhaps impurities or boundary or thickness effects.
We  find analytical expressions for the critical exponents of the general model and show that they satisfy the Rushbrooke identity.
An important subclass of models exhibit second order phase transitions. A study of the specific heat shows that 
 general models can also describe holographic superconductors undergoing first, second and third (or higher) order phase transitions. We discuss how small
deformations of the HHH model can lead to the appearance of resonance peaks in the conductivity, which increase in number and 
become narrower as the temperature is gradually
decreased, without the need for tuning mass of the scalar to be close to the Breitenlohner-Freedman bound.
Finally, we investigate the inclusion of a generalized ``theta term" producing Hall effect without magnetic field. 

\end{abstract}

\end{center}

\clearpage

\tableofcontents

\section{Introduction}

The AdS/CFT correspondence \cite{malda,gubser,witten} has been recently applied to provide a gravitational description of systems that undergo superfluid or superconducting phase transitions.
Holographic superconductors \cite{Gubser:2008px,Hartnoll:2008vx,Hartnoll:2008kx}
reproduce various features of the physics of BCS superconductors, and they represent a promising avenue
for constructing tractable models of strongly interacting systems like high $T_c$ superconductors or other systems exhibiting non-Fermi liquid behavior.

The basic model of \cite{Hartnoll:2008vx} (henceforth, the HHH model) depends only on one free parameter and obviously does not have enough room to incorporate
non-universal features of the diverse physics that one can find in real superconductors \cite{Parks}. 
Clearly, it is of interest to construct and investigate more general frameworks for holographic superconducting systems.
Simple generalizations of the HHH model have been proposed in \cite{Franco:2009yz,Franco:2009if,Aprile:2009ai}, by introducing arbitrary couplings and an arbitrary
potential which are functions of the complex scalar field.
A convenient setup is to replace the complex scalar field by a neutral scalar field $\eta $ and  a  St\" uckelberg field, and
then write the most general gauge-invariant Lagrangian which involves general functions of the real scalar field $\eta $.
In this paper we investigate the main properties of these models (reviewed in section 2).
We also discuss a further generalization obtained by introducing of a generalized  ``theta term" of the form $\Theta (\eta )\ F\wedge F$.

 In weakly-coupled superconductors, Anderson's theorem (see \textit{e.g.} \cite{Parks, GdG}) implies that the thermodynamic properties of any superconductor remain unchanged if
a perturbation does not break time-reversal invariance and does not cause a long-range spatial variation of the order parameter.
This basically implies that non-magnetic impurities do not affect the thermodynamic properties of a superconductor.
There are many examples of violation of this theorem in real superconductors, which is itself a sign that the underlying physics involves strong coupling.
But in general one expects that  time-reversal breaking perturbations will give rise to new  effects in the system. 
It is therefore of interest to study mechanisms for spontaneous (or explicit) breaking of time-reversal symmetry.
Within the context of our general models, a natural way to introduce a time-reversal violating perturbation is in terms
of a bulk term of the form $\Theta(\eta )\ F\wedge F$. Choosing $\Theta(\eta )$ such that $\Theta(0)=0$, this term will only be turned on in the condensed phase.
While this term does not affect the temperature dependence of the order parameter,
we will find that it has striking effects in the conductivity.

This paper is organized as follows. In section 2, we present our general model and derive the equations of motion. Section 3 contains analytic results for the behavior of condensate near the phase transition, including critical exponents. Section 4 is devoted to the study of the  free energy and specific heat near the critical point.
Section 5 studies the effects of deformations on the conductivity. 
Section 6 investigates Hall conductivities arising from the generalized theta term. A discussion is given in  section 7. 

\bigskip
\bigskip

\noindent {\bf Note added:} As this paper was being finalized, \cite{Herzog:2010vz} appeared, with  some overlap with section 3.2 concerning analytic expressions
for critical exponents.

\section{General holographic superconductors}
\def\p{\partial }
\def\vphi{\varphi}
\def\s{\sigma }

\label{section_general_HS}

We are interested in studying a system undergoing a phase transition where a $U(1)$ symmetry is spontaneously broken at a certain critical temperature $T_c$ through holographic methods. On general grounds  \cite{Klebanov:1999tb}, the global $U(1)$ current is dual to a gauge field in AdS. Spontaneous symmetry breaking in the boundary theory is dual to the Higgs mechanism in the bulk. It is then natural to consider the most generic theory under such premise. Extending previous proposals in \cite{Franco:2009yz,Franco:2009if}, such generic models have been considered in \cite{Aprile:2009ai}. They are described by the 
following $U(1)$ invariant $3+1$ dimensional\footnote{This model can  be generalized to arbitrary $d+1$ dimensions \cite{Aprile:2009ai}. Here we focus on the case $d=3$ for concreteness.} action
\be
S= {1\over 16\pi G_N}\int d^{3+1}x \sqrt{-\hat g}\left( R -{1\over 4} G(\eta)\ F^{\mu\nu} F_{\mu\nu} +{6\over L^2} U(\eta) -{1\over 2} 
(\partial \eta)^2 -{1\over 2} J(\eta)\, (\partial_\mu \theta -A_\mu )^2\right)\ .
\label{cero}
\ee
The Lagrangian describes the dynamics of a metric $\hat g_{\mu\nu}$, a $U(1)$ gauge field $A_\mu$, a real scalar  field
$\eta$ and a St\" uckelberg field $\theta $. 
We work with dimensionless fields $A_\mu$, $\eta $, while $[\theta]=-1$ (note that $[J(\eta)]=2$). The canonical normalization for the kinetic terms
can be restored by an appropriate rescaling involving the Newton constant $G_N $.
The action can also be cast in terms of a complex scalar field
\be
S= {1\over 16\pi G_N}\int d^{3+1}x \sqrt{-\hat g}\left( R -{1\over 4} G(\bar \psi\psi )\ F^{\mu\nu} F_{\mu\nu} +{6\over L^2} U(\bar\psi\psi ) -
K(\bar\psi \psi)\, D^\mu \bar\psi D_\mu \psi \right)\ ,\qquad 
\label{cerodos}
\ee
where
\be
\psi =\hat \eta\ e^{iq\, \theta} \ ,\qquad {d\hat\eta \over\hat \eta } = {|q|\ d\eta\over \sqrt{J(\eta) }}\ ,\qquad K\equiv {J\over 2q^2\bar\psi\psi}\ ,\qquad D_\mu =\p_\mu - i q A_\mu \ .
\ee
The two models (\ref{cero}) and (\ref{cerodos}) are not equivalent in some cases, since (\ref{cero}) permits negative values for the scalar field, whereas $\bar\psi\psi$, appearing in the couplings,
is positive. In particular, they are equivalent when  $G(\eta)$, $U(\eta)$ and $J(\eta)$ depend only on even powers of $\eta$ with $J=q^2\eta^2+O(\eta^4)$.

The models typically describe holographic superconducting phase transitions in 2+1 dimensional systems. The phase transitions were
shown to be either continuous or discontinuous, depending on the choice of 
the $G(\eta)$, $U(\eta)$ and $J(\eta)$ couplings \cite{Franco:2009yz,Franco:2009if,Aprile:2009ai}. As we will see later on, the continuous phase transitions can be of arbitrary order in Ehrenfest's classification. Interestingly, the string and M-theory realizations of holographic superconductors \cite{Gubser:2009qm,Gauntlett:2009dn} belong to this general class of models (\ref{cero}).

The model (\ref{cero})  admits yet one more generalization (not studied in  \cite{Aprile:2009ai}) given by
\be
\delta S =\int d^{3+1}x \ {1\over 4}\ \Theta (\eta ) \ \epsilon^{\mu\nu\rho\sigma}F_{\mu\nu}F_{\rho\sigma}\ .
\label{tet}
\ee
For generic $\Theta (\eta )$, this term violates parity and time-reversal symmetry. However,
for special couplings of the form $\Theta = \theta_0 \eta^{2k+1}$, with integer $k$, parity and time-reversal symmetries still hold if we assume that
$\eta $ is a pseudoscalar transforming as $\eta\to -\eta $ under $P$ and $T$. Note that this parity-preserving option cannot be implemented in the formulation (\ref{cerodos}) in terms of a complex scalar field  (in particular,   in the model of \cite{Hartnoll:2008vx}).
Similar terms appear from M-theory compactifications, in consistent Kaluza-Klein truncations that include a complex scalar field and a real scalar field \cite{Gauntlett:2009bh}.
In this case, $\Theta $ is a specific function of the real scalar field.

For reference purposes, it is useful to define two particular models with couplings having the following behavior at small $\eta $:
\be
{\rm Model\ I:}\qquad G(\eta )\cong 1+\kappa \eta^2 +O(\eta^4) \ ,\qquad U(\eta ) \cong 1 + {1\over 6}\ \eta^2+O(\eta^4)\ ,\qquad J(\eta ) \cong q^2 \eta^2+O(\eta^4)\ .
\label{stm}
\ee
with $\Theta (\eta )=0$. 
The HHH model \cite{Hartnoll:2008vx} is the particular case in which $\kappa =0$, with $U= 1 + {1\over 6}\ \eta^2$ and $J=q^2 \eta^2$; the  
string and M-theory models \cite{Gauntlett:2009bh, Gubser:2009qm,Gauntlett:2009dn} also   belong to this class (see \cite{Aprile:2009ai}).
It will be shown that all these models have mean field critical exponents. Nonetheless, the physics described by models I with different couplings will be shown
to exhibit strikingly different features.
A slight generalization of Model I is to allow for odd powers of $\eta $.
We will show that including a power $\eta^3$ in the expansion of $G$, $U$ or $J$ already modifies the critical exponents.

The second class of models that we will consider
allows for non-analytic terms in the small $\eta $ expansion:
\be
{\rm Model\ II:}\ \ \  G(\eta )\cong 1+\kappa \eta^2 +g_0 |\eta|^{a}\ ,
\ \ \  U(\eta ) \cong 1 + {1\over 6}\ \eta^2 + u_0 |\eta|^b\  ,\ \ \ 
J(\eta ) \cong q^2 \eta^2+j_0|\eta|^c \ .
\label{stmm}
\ee
Here $a$, $b$, $c$ are assumed to be real, positive numbers, with $a,\ b,\ c>2$ ($G(\eta)$ and $J(\eta)$ must be positive definite for unitarity and $u_0<0$ for stability). 
{}For generic values of $a,b,c$, this model contains non-analytic interactions in $\eta $. Since the classical solutions we will study have $\eta \geq 0$,\footnote{In the formulation in (\ref{cerodos}) in terms  of a complex scalar field  the couplings depend on $\bar\psi\psi $, which is always positive.} the non-analytic dependence on $\eta $ is
irrelevant at the classical level. Higher order terms can be added to the expansion. We will see that such terms do not modify the dynamics close the phase transition, but can be important in other regimes, such as low temperatures.

A problem of general interest is to determine whether real materials can be approximated by general holographic models to some extent. 
If this is the case, the results in this paper can be used to determine the best holographic fit, within our class of models, for a given material. 

Before moving on, we would like to emphasize some important points regarding the interpretation of models with non-analytic interactions, such as type II models (\ref{stmm}) in the case of generic (non-integer) exponents $a,b,\ c$.
It will be shown below that the presence of non-analytic terms in model II allows critical exponents to be tuned to arbitrary values in a certain range, thus
including values away from the standard Landau theory. However, we recall that one does not expect non-mean field theory behavior at large $N_c$. 
This is consistent with the fact that one does not expect any non-analytic terms in the potential in {\it classical} Lagrangians originating from string/M theory compactifications,  in general.
Non-analytic terms might effectively be induced by quantum corrections but, since in string theory quantum corrections are controlled by $1/N_c$, 
the coefficients of such non-analytic terms would be small in the large $N_c$ approximation, vanishing in the classical supergravity limit.
Finally, we stress that (as in most approaches to holographic superconductors) here we will not be concerned about the UV completeness of the model. 
Type II models (\ref{stmm}) with generic exponents may  just be regarded as a phenomenological approach that can capture some 
specific features of real superconductors, in particular, to incorporate/parametrize certain interesting behaviors close to the phase transition.    

\subsection{Ansatz and equations of motion: General setup}

Upon fixing the gauge $\theta =0$, the action (\ref{cero}) takes the  form
\be
S= {1\over 16\pi G_N}\int d^{3+1}x \sqrt{-\hat g}\left( \ R -{1\over 4} G(\eta)\ F^{\mu\nu} F_{\mu\nu} +{6\over L^2} U(\eta) -{1\over 2} (\partial \eta)^2 -{1\over 2} J(\eta) A_\mu A^\mu\right)\ .
\label{uno}
\ee
We now consider the following ansatz
\be
ds^2= -g(r) e^{-\chi (r)} dt^2 +{dr^2\over g(r)} + r^2(dx^2+dy^2)\ ,\qquad A=\phi(r)dt\ ,\qquad \eta =\eta (r)\ .
\label{solu}
\ee
The effective Lagrangian takes the form
\be\label{L1}
\sqrt{-\hat g} {\cal L} = -2 e^{-{\chi\over 2}} (r g)'+ {r^2\over 2} G(\eta ) e^{\chi\over 2} {\phi'}^2+{6r^2\over L^2}  e^{-{\chi\over 2}} U(\eta ) -{r^2\over 2} e^{-{\chi\over 2}} g {\eta'}^2
+{r^2\over 2g} e^{\chi \over 2} J(\eta ) \phi^2.
\ee
It follows that the equations of motion reduce to
\bea
&&\chi'+ {r\over 2} {\eta'}^2+ {r\over 2g^2} e^\chi J(\eta)\phi^2=0\ ,
\label{Eq1}\\
&& {1\over 4}\ {\eta'}^2+ {G(\eta)\over 4g}\ e^\chi {\phi'}^2+{g'\over rg} +{1\over r^2} -{3\over L^2 g}\ U(\eta) +{1\over 4g^2}\ e^\chi J(\eta) \phi^2=0\ ,
\label{Eq2}\\
&& \phi''+\phi' \ \left( {2\over r} +{\chi'\over 2} +{\p_\eta G \eta'\over G} \right)- {J(\eta)\over g G(\eta)} \ \phi =0\ ,
\label{Eq3}\\
&& \eta'' +\eta'  \ \left( {2\over r} -{\chi'\over 2} +{g'\over g} \right)+ {1\over 2g}\ e^\chi \p_\eta G \ {\phi'}^2 + {6\over L^2g}\ \p_\eta U + {1\over 2g^2}\ e^\chi \p_\eta J\ {\phi}^2 =0\ .
\label{Eq4}
\eea

\medskip

In the limit $q\rightarrow \infty$ keeping $q\eta$ and $q\phi$ fixed, the matter source drops out from the Einstein equations (\ref{Eq1}), (\ref{Eq2}) and the solution is just the AdS-Schwarzschild black hole,
\bea
g(r)=r^2 \Big(1-\frac{r_h^3}{r^3}\Big),\qquad\qquad\chi(r)=0\ ,
\label{geres}
\eea
where $r_h$ represents the position of the horizon. In this limit the scalar and Maxwell equations keep the same structure,
\bea
&& \phi''+\phi' \ \left( {2\over r} +{\p_\eta G \eta'\over G} \right)- {J(\eta)\over g G(\eta)} \ \phi =0\ ,
\label{Eq3uno}\\
&& \eta'' +\eta'  \ \left( {2\over r} +{g'\over g} \right)+ {1\over 2g}\ \p_\eta G \ {\phi'}^2 + {6\over L^2g}\ \p_\eta U + {1\over 2g^2}\ \p_\eta J\ {\phi}^2 =0\ .
\label{Eq4uno}
\eea
In \cite{Aprile:2009ai} (generalizing the discussion of \cite{Hartnoll:2008kx}) it was found that, even for small values of $q$, backreaction does not alter the thermodynamic properties of the system significantly.
Therefore, in what follows we will use this no backreaction approximation, with the purpose of simplifying the numeric analysis and obtaining some analytical results.

It is convenient to introduce a new coordinate $z=r_h/r$,  so that the horizon is at $z=1$ and the boundary is located at $z=0$. The equations of motion become
\bea
&& \p_z^2\vphi+ \p_z\vphi \ \left({\p_\eta G \p_z\eta \over G} \right)- \frac{1}{z^2(1-z^3)}{J(\eta)\over G(\eta)} \ \vphi =0\ ,
\label{Eq3tres}\\
&& \p_z^2\eta - \p_z\eta  \ \frac{(2+z^3)}{z(1-z^3)}+ \frac{z^2}{2(1-z^3)}\p_\eta G \ (\p_z \vphi )^2 + {6\over L^2}\frac{1}{z^2(1-z^3)}\ \p_\eta U + 
{\p_\eta J\over 2(1-z^3)^2}\ {\vphi}^2 =0\ ,
\label{Eq4tres}
\eea
where we have defined $\vphi \equiv \phi/r_h$.
In these new variables the asymptotic behavior of $\eta$ and $\vphi$ is
\bea\label{asymp}
\eta_\infty=
\frac{\eta^{(1)}}{r_h}z+\frac{\eta^{(2)}}{r_h^2}z^2,\qquad\qquad
\vphi_\infty=
\frac{\mu}{r_h}-\frac{\rho}{r_h^2}z\ .
\eea
Note that the asymptotic behavior of $\eta$  --in turn related to the conformal dimension of the condensing operator $O_i$-- is due to  the special choice of  mass in (\ref{stm}), (\ref{stmm}),
which is also the choice of the HHH model. The behavior for different masses has been studied in \cite{Horowitz:2008bn, Denef:2009tp}.

If $\eta^{(1)}=\eta^{(2)}=0$, then the solution is $\eta=0$ in the whole space.
As explained in \cite{Hartnoll:2008vx,Hartnoll:2008kx}, there are two schemes that describe spontaneous symmetry breaking. One corresponds to setting $\eta^{(1)}=0$, such that a global $U(1)$ symmetry in the boundary field theory is broken spontaneously
by a condensate $\langle O_2\rangle =\eta^{(2)}\neq 0$. The other scheme corresponds to $\eta^{(2)}=0$; in this case the superconducting regime is described by
a condensate $\langle O_1\rangle =\eta^{(1)}\neq 0$.

\subsection{General models, CFT deformations and quantum critical points}

\label{section_deformations}

Let us conclude this section with a few general remarks. We are interested in continuous phase transitions where a $U(1)$ symmetry is spontaneously broken. At the phase transition, correlation lengths diverge and the physics is expected to be dominated by a (strongly interacting) conformal fixed point. Universality then suggests that few parameters are necessary to distinguish the fixed point. In our case, since $\eta $ is small near the critical point, we expect the identity of such fixed point to be controlled by the lowest order terms in the small condensate expansion of the functions $G(\eta)$, $U(\eta)$ and $J(\eta)$ in (\ref{stmm}). More concretely, the dimension and charge of the CFT operators are given by the coefficients of the quadratic terms. In addition, as we will see below, critical exponents are dictated by the exponents of the subleading terms. It is then clear that our general models have plenty of room for varying the functions without 
changing this data, by changing the coefficients of subleading terms (as we will do in section \ref{section_conductivity}) or including higher order terms. Some of the perturbations, like modifying the potential function $U(\eta)$ are rather straightforward to implement. These higher order corrections can be interpreted as perturbations of the conformal fixed point by irrelevant operators. As such, they would not alter the properties of the conformal fixed point, but would be in turn relevant away from it when looking at other dynamical properties, among which perhaps the most natural one is the conductivity. From this point of view, it is natural to expect that these general models might provide a framework for studying quantum critical points, namely second order phase transitions at zero temperature as a function of variable couplings.\footnote{ A recent discussion on quantum phase transitions in the holographic context can be found in \cite{Iqbal:2010eh}.}

In general, subleading terms in the $G(\eta)$, $U(\eta)$ and $J(\eta)$ functions represent interactions that might effectively account for
the result of integrating out massive modes, giving rise to a low-energy effective action for $\eta$. From the dual field theory  viewpoint, this would translate into an effective free energy functional for the low energy effective degrees of freedom, namely the operator $\langle O_i\rangle$ dual to the $\eta$ field. This free energy functional will be discussed in section 4. 
As discussed earlier, non-mean field critical exponents 
arise if one includes non-analytic  
 subleading terms in $G(\eta)$, $U(\eta)$ and $J(\eta)$, and such terms should be suppressed in the large $N_c$ limit.
Alternatively, one may  take a phenomenological approach and use the parameters of the model to fit properties of real systems, possibly allowing them to be $\mathcal{O}(1)$.

\section{Behavior near the phase transition: temperature dependence of the order parameter and critical temperature}

As shown in  \cite{Franco:2009yz,Franco:2009if,Aprile:2009ai}, the models above exhibit phase transitions at a certain critical temperature $T_c$. Depending on the choice of parameters, these transitions have been shown to be qualitatively different; ranging from first order to second order with different critical exponents.\footnote{Here in what follows, we use the term second order to refer to continuous phase transitions. In section \ref{section_free_energy_holographic}, we discuss the behavior of the free energy across the transition in more detail.} In the remainder, we will concentrate on second order transitions. Previous studies performed in the past have been mainly numeric. On the other hand, it would be desirable to have an analytic handle on crucial properties characterizing the transition. In this section we describe a simple method to determine the critical temperature and an analytic expression of the temperature dependence of the condensate close to the phase transition, including critical exponents.

\subsection{Critical temperature}

A convenient way to determine the critical temperature of second (or higher)-order phase transitions
is as follows. Near the critical point, $\eta\to 0$, so we can write a perturbative expansion
\be
\label{eq_for_Tc}
\eta =\epsilon \tilde \eta + O(\eps^2) \ ,\qquad \varphi =\varphi_0 +\epsilon^2 \tilde \varphi +O(\epsilon^3)\ ,
\ee
where $\varphi_0$ and $\tilde \eta$ satisfy the equations
\bea
&& \p_z^2\vphi_0=0 ,\\
&&\p_z^2\tilde\eta -\p_z \tilde\eta  \ \frac{(2+z^3)}{z(1-z^3)}+ \frac{z^2}{2(1-z^3)}\p_{\tilde\eta}G \ (\p_z \vphi_0)^2 + {6\over L^2}\frac{1}{z^2(1-z^3)}\ \p_{\tilde\eta} U + 
{\p_{\tilde\eta} J\over 2(1-z^3)^2}\ {\vphi_0}^2 =0\ .
\label{etaeq}
\eea
To leading order in $\epsilon $, only the quadratic terms in $G(\eta)$, $U(\eta)$ and $J(\eta)$ contribute to the $\tilde \eta $ equation.
Thus in what follows we keep only the quadratic terms of  model I or model II, and the equation for $\tilde \eta $ becomes linear.
The solution for $\varphi_0$ is simply
\be
\varphi_0 = {\mu \over r_h}\ (1-z)={\rho\over r_h}\  (\frac{1}{r_h}-\frac{z}{r_h})\ .
\ee
Substituting the expansion (\ref{stm}) into (\ref{etaeq}), we find 
that the differential equation (\ref{etaeq}) depends only the combinations $\tilde \kappa \equiv\kappa/q^2$ and $A \equiv\mu q/r_h$.
The boundary conditions for the $\eta $ equations are 
\be
\tilde \eta(1) = \eta_h\ ,\qquad \tilde \eta'(1) = {1\over 3} \eta_h {\mu^2\over r_h^2} \ (\kappa -q^2)=  {1\over 3} \eta_h A^2 \ (\tilde \kappa -1)
\ .
\ee 
Since the $\eta $ equation is linear, the solution is of the form
\be
\tilde\eta = \eta_h F(A,\tilde \kappa ; \ z)\ .
\ee
Expanding near $z=0$, we have
\be
{\tilde \eta \over \eta_h} = c_1(A,\tilde \kappa )\ z +c_2(A,\tilde \kappa )\ z^2+ ....
\ee
In the scheme where $\langle O_2\rangle$ is non-vanishing, the critical temperature is then determined from the requirement that 
\be
c_1(A, \tilde \kappa ) =0\ .
\ee
In the scheme where $\langle O_1\rangle$ is non-vanishing, the requirement is
\be
c_2(A,\tilde \kappa ) =0\ .
\ee
In either scheme, these conditions determine $A=A(\tilde \kappa )$. Then one uses $T= 3r_h/4\pi$ and $r_h = \mu q/A $ or $r_h=\sqrt{\rho q/A}$.
This gives a critical temperature of the general form
\be
T_c = {3\over 4\pi}\ q \mu f(\kappa /q^2)\ ,
\label{mumu}
\ee
or
\be
T_c ={3\over 4\pi}\ \sqrt{q \rho} \ \sqrt{f(\kappa /q^2)}\ ,
\label{rhorho}
\ee
which are appropriate for calculations at fixed $\mu $ or fixed $\rho $, respectively. Here $f\equiv 1/\sqrt{A(\tilde \kappa) }$.

\begin{figure}[tbh]
\centering
{\includegraphics[width=7.5cm]{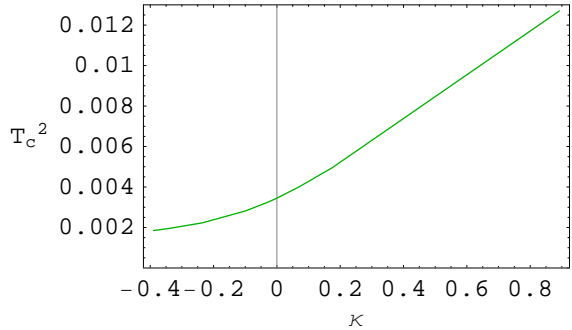}}
\caption{$T_c^2$ vs. $\kappa  $ in second (or higher) order phase transitions in Model I or Model II ($q=\mu=1$).}
\label{figspec1}
\end{figure}

Since the method implies solving an ordinary second order differential equation, the critical temperature can be determined with high accuracy.
Here we use the scheme where $\langle O_2\rangle$ is non-vanishing.
In particular, for fixed $\mu =1$, we find the value  $T_c(\kappa =0)=0.05874734 \ q$ (or $T_c(\kappa =0)= 0.1184267 \ \sqrt{q}$ if we work at fix charge density $\rho=1$, in agreement
with \cite{Hartnoll:2008vx}). A plot of $T_c^2$ as a function of $\kappa /q^2$ at fixed $q$ is shown in figure \ref{figspec1}.
We see that at large $\kappa  $, $T_c^2$ becomes linear with $\kappa  $ and, for large negative $\kappa  $, $T_c^2 $ tends to zero monotonically.

Thus, the parameter $\kappa $ can be used to tune the critical temperature to any desired value.
A similar role is played by the parameter $q$ or the scalar  (which for our models was set to the special value $m^2=-2$), i.e. they 
also affect the critical temperature. 

\subsection{Critical curve for the order parameter}

As numerically shown in \cite{Franco:2009yz,Franco:2009if,Aprile:2009ai}, the model exhibits second order phase transitions in certain regions of its parameter space. In second order phase transitions the order parameter continuously approaches zero as we approach the critical temperature from below. Translated into our holographic approach, close to the phase transition the scalar $\eta$ should be arbitrarily small. A useful approach for studying the vicinity of the phase transition is then to solve the equations of motion (\ref{Eq3tres}) and (\ref{Eq4tres}) in terms of a series expansion of $\eta$ and $\vphi$ near the horizon, located at $z=1$.\footnote{An alternative method that leads to the same result follows the idea of the previous subsection, namely to look in the vicinity of $T_c$ and solve the differential equations in perturbation theory in powers of $\eta $ to the next order.}

\bea\label{serie1}
\eta^{(N)}(z)&=&\eta_h+\eta_1(1-z)+\eta_2(1-z)^2 +\ldots  +\eta_N (1-z)^N \ ,
\nn\\
\vphi^{(N)}(z)&=&\vphi_1(1-z)+\vphi_2(1-z)^2+\ldots + \vphi_N (1-z)^N \ ,
\label{solution_expansion}
\eea
where we truncate the series at some order $N$.

The coefficients $\eta_j$ and $\vphi_j$ can be solved in terms of $\eta_h$ and $\vphi_1$ using the equations of motion for every $j>1$. The series 
(\ref{serie1}) converges to the solutions $\eta(z)$ and $\vphi(z)$ as $N\to \infty$. Then, we match these expansions with the asymptotic behavior at the boundary (\ref{asymp})
\bea
\left\{
\begin{array}{c}
\eta^{(N)}(0)=\eta_{\infty}(0)\\\\
\p_z\eta^{(N)}(0)=\p_z\eta_{\infty}(0)
\end{array}\right.\qquad\qquad
\left\{\begin{array}{c}
\vphi^{(N)}(0)=\vphi_{\infty}(0)\\\\
\p_z\vphi^{(N)}(0)=\p_z\vphi_{\infty}(0)
\end{array}\right.
\eea

We consider model II, which has eight parameters: $\kappa$, $q$, $\ g_0$, $u_0$, $j_0$, $a$, $b$ and $c$. As we will see, model II captures all the terms that are important close to the phase transition for a completely generic model. For concreteness, we focus on the $\langle O_2\rangle = 0$ scheme. There are five parameters $\eta_h,\vphi_1$ (from the horizon side) and $\eta^{(1)},\ \mu,\ \rho$ (from the boundary side). These conditions can be solved for four parameters in terms of either $\rho$ or  $\mu$.\footnote{Another possibility is to implement the matching at some intermediate point $z_m$, with $0<z_m<1$, as done in \cite{Gregory:2009fj}. While having variable $z_m$ allows one to obtain better solutions at fixed order $N$ in the expansion, it introduces complicated algebraic equations that prevent the application of this method at large $N_c$.} Furthermore, close to the critical temperature, $O_1 =\eta^{(1)}$ is linear in $\eta_h$.

Expressing $r_h$ in terms of $T$ using $r_h=4\pi T/3$, we find the following generic structure
\bea\label{MasterEq}
1- \frac{T}{T_c^{(N)}(\kappa ,q^2)} = A_N(\kappa ,q^2)\langle O_{1}\rangle ^2+g_0 B_N(a) \langle O_{1}\rangle ^{a-2}+u_0C_N(b) \langle O_{1}\rangle ^{b-2}+j_0 D_N(c)\langle O_{1}\rangle ^{c-2}+ \ldots
\eea
where ``$\ldots$" denotes terms that are of higher order in $\eta_h$. The key point is that the structure of (\ref{MasterEq}) remains the same at each successive order in $N$. 
The different functions (including $T_c^{(N)}$) get corrections, but the general form of the equation, which remains unchanged, already reveals
interesting information on the phase transition. In particular, it gives the explicit functional dependence of the order parameter on the temperature, hence
it gives  the explicit analytic expression for the critical exponent $\beta $. For  $a,\ b,\ c \geq 4$, the leading term  in  (\ref{MasterEq}) is $\eta^2$. In this case we find
\be
\langle O_1\rangle \cong {\rm const.}\ \left( 1-{T\over T_c} \right)^{1\over 2}  \ ,\qquad {\rm for}\ T\cong T_c\ ,
\ee 
as in mean field theory. When either  $a$, $b$ or $c$ is less than $4$, then the leading term is $\eta^{a_0 -2} $, where  $ a_0\equiv {\rm Min}\{a,\ b,\ c\}$.
 In this case, we find
\be
\langle O_1\rangle \cong {\rm const.}\ \left( 1-{T\over T_c} \right)^\beta \ ,\qquad \beta ={1\over a_0-2}\ ,\ \ a_0 = {\rm Min}\{a,\ b,\ c\}\ .
\label{critical_exponent_condensate}
\ee 
Because of the presence of the term $A_N(\kappa ,q^2)\langle O_1\rangle ^2$, generically one has the bound $\beta \geq 1/2$.
This value, as well as the possibility of deriving it from an expansion like (\ref{solution_expansion}), was already advanced in \cite{Franco:2009if,Aprile:2009ai} and is in perfect agreement with the numeric results. Remarkably, the correct value of exponents can be determined from the expansion at lowest order, in contrast to quantities such as the critical temperature. Their values are robust provided the corresponding coefficient in (\ref{MasterEq}) does not go to zero in the $N\to \infty$ limit, something that can be easily checked. Recently, the same value of the critical exponent $\beta $ was independently derived in \cite{Herzog:2010vz}, where it was discussed in the context of general $U(\eta)$, $J(\eta)$.
{} Furthermore, in the specific case in which the Breitenlohner-Freedman (BF) bound is saturated, \cite{Herzog:2010vz} found an analytic solution from which the numerical coefficients in (\ref{MasterEq}) can be determined. In section \ref{section_free_energy_holographic}, we explain how the same result can be deduced from minimizing the free energy.

In principle, it is possible that $A_{N\to\infty}(\kappa ,q^2)$ vanishes for some real value of $\kappa/q^2$. If this is the case, one might think that
in this special point of parameter space one can obtain models with $\beta <1/2$ by taking
$a,b,c> 4$. However, numerical study suggests that before this occurs the transition changes into a first order phase transition.
Finally, let us mention that an equation with the same structure as (\ref{MasterEq}) is obtained in the  $\eta^{(1)}=0$ scheme where $\langle O_2\rangle $ takes an expectation value, so the same considerations apply in this case.

An important feature we learn from the critical curve (\ref{MasterEq}) is that $T_c$ depends only on $\kappa $ and $q$, i.e. on the quadratic terms
in the expansion of $G(\eta)$, $U(\eta)$ and $J(\eta)$. By introducing a general mass $m^2$ parameter in $U$, one sees that $T_c$ is more generally a function of $\kappa $, $m$, and $q$ 
(the mass dependence of $T_c$ was previously discussed in detail in \cite{Denef:2009tp}). This important feature is confirmed by 
the analysis of the previous subsection, where we saw that the critical temperature
is determined by a linear differential equation and it is thus unaffected by 
higher order terms in  $G(\eta)$, $U(\eta)$ and $J(\eta)$. In particular, this feature was the content of conjecture 2 of \cite{Aprile:2009ai}.

\section{Free energy}

\label{section_free_energy_holographic}

Let us now turn to the study of the free energy for our general models. As standard, the free energy is given by the gravity on-shell action. This quantity is divergent, and thus needs to be regularized before further physical interpretation. Regularizing the action with a radial cut-off $r_B$, it is not difficult to show that the structure of such action is

\begin{equation}
S^{(1, \, \mu)}=\frac{\mu\, \rho}{2} + \frac{r_B\, (\eta^{(1)})^2}{2\, L} + \cdots \, ,
\end{equation}
where the dots stand for finite terms irrelevant at this stage. In addition, we keep the superscripts to remind that we will focus on the case with fixed chemical potential and the $\langle O_2\rangle =0$ scheme. Since this action is divergent, we need counterterms to holographically renormalize it.  Given the structure of the on-shell action, it is nevertheless clear that the same counterterm as in \cite{Franco:2009yz} will do the job

\begin{equation}
\Delta S_{B}=-\frac{1}{2}\, \int_{\rm B}\, [\sqrt{\gamma}\, \eta^ 2]
\end{equation}
being $\gamma$ the induced metric on the boundary $B$. Then, the free energy for the general models (\ref{cero}) takes the  form 
\beq
\label{Wooo}
W=-\frac{\mu\, \rho}{2}-\frac{O_1\, O_2}{2}+\frac{r_H^3}{2} \int\ dz\  \left[
\frac{\eta}{2}\p_{\eta}G\, (\p_z\varphi )^2+\eta\p_\eta J \, \frac{\varphi^2}{2z^2(1-z^3)}+ \frac{1}{z^4}(\eta\p_{\eta}U-2U),
\right]\ .
\eeq
Near the critical point, where $\eta $ is small, we can use the expansion for 
 type II models (\ref{stmm}) and write
\beq
\label{W}
W=-\frac{\mu\, \rho}{2}-\frac{O_1\, O_2}{2}+\frac{r_H^3}{2}\int dz\, \left[ 
\frac{a\, g_0}{2}\, z^a\, \chi^a\, (\p_z \varphi )^2 + \Big(q^2\chi^2+\frac{m\, j_0}{2}\, z^{m-2}\,\chi^m\Big) \, \frac{\varphi^2}{(1-z^3)}+ \frac{u_0\, (b-2)}{z^{4-b}}\, \chi^b \right] \ .
\eeq
where we have introduced $\chi=z^{-1}\, \eta$ and dropped the superscript.  We have also set $\kappa=0$ to maintain the same critical temperature.

In order to further proceed, we can use the series solution to the equations of motion as in (\ref{serie1}). Inserting the resulting expressions in (\ref{W}) we obtain 
\begin{eqnarray}
W^{(N)} &= & W_0^{(N)}(T)(\kappa,q^2,T) + W_2^{(N)}(\kappa,q^2,T) O_1^2 + g_0 \, W_c^{(N)} (\kappa,q^2,T,a)\, O_1^a  \nonumber \\ 
& + & u_0 \, W_b^{(N)} (\kappa,q^2,T,b)\, O_1^b + j_0 \, W_c^{(N)} (\kappa,q^2,T,c)\, O_1^c + \ldots
\label{master_free_energy}
\end{eqnarray}
where ``$\ldots$" denotes terms with higher powers of $\eta_h$. We keep this schematic form since, once again, the particular expressions of the coefficients $W_i^{(N)}$ are cumbersome and not too illuminating. As for (\ref{MasterEq}), the general structure of (\ref{W}) is independent of $N$. Indeed, an identical discussion applies also here. We have also verified that the expansion in (\ref{master_free_energy}) approximates the exact (numerical) free energy close to the phase transition.

The critical temperature at order $N$ corresponds to the value at which $W_2^{(N)}$ changes its sign
\beq
W_2^{(N)} \sim (T-T_c^{(N)}(\kappa,q^2)) + \ldots
\eeq
$W_0^{(N)}$ and $ W_c^{(N)}$ remain finite at $T_c^{(N)}(\kappa,q^2)$.

As discussed in section 2.2, the holographic free energy functional (\ref{W}) can be thought of as a sort of generalized version of the Landau-Ginzburg free energy. For integer values of the exponents $a$, $b$ and $c$, we can think that the extra couplings in $G(\eta)$, $U(\eta)$ and $J(\eta)$ correspond to the effect of higher dimensional operators in the bulk effective field theory. Such terms are known to arise in consistent truncations of String/M-theory \cite{Gubser:2009qm,Gauntlett:2009dn}. Further exploring their possible origin in the boundary field theory is of great interest, but beyond the scope of this paper.

\subsection{Specific heat and higher-order phase transitions}

\begin{figure}[tbh]
\centering
{\includegraphics[width=7.5cm]{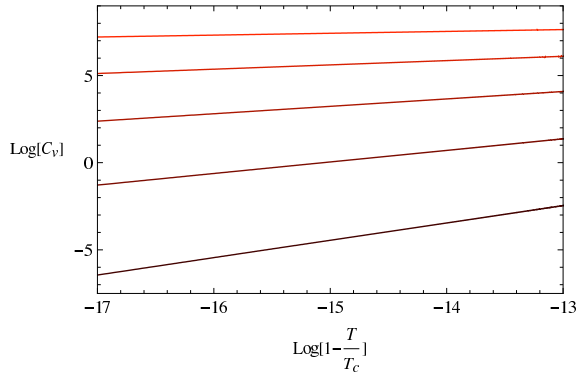}}
\caption{$\Delta c_v$ as a function of temperature close to the phase transition for $J=\eta^2 - \eta^c + \eta^4$, with $c=3+k/5$, $k=0,\ldots,4$.}
\label{Cv_T}
\end{figure}

The free energy (\ref{W}) contains all the relevant information about the system close to the phase transition. By minimizing it with respect of $O_1$ we can obtain the physical VEV of the condensate as a function of the temperature. In particular, we can consistently check that the critical exponent $\beta$ obtained in this way coincides with the one we got in the previous section. Substituting such value into the free energy (\ref{master_free_energy}), we can determine the behavior of the specific heat close to the critical temperature. We obtain 
\be
\Delta c_v = -T {d^2 W\over d T^2} \equiv {\rm const.} (T_c-T)^{-\alpha} \ ,\ \ \ \ \ \ \alpha=-{4-a_0 \over a_0 -2} \ .
\label{critical_exponent_specific_heat}
\ee
where $a_0={\rm Min}\{a,\ b,\ c\}$ appeared earlier in (\ref{critical_exponent_condensate}). Note in particular that we automatically get $\alpha+2\beta=1$. This result was also recently derived in \cite{Herzog:2010vz} for the case with non-trivial $U(\eta)$ and $J(\eta)$, simply using the scaling relations. Our result holds more generally and, since our derivation is independent of the scaling relations, it provides an independent proof that they are satisfied as we discuss below. Figure \ref{Cv_T} shows $\Delta c_v$ as a function of temperature close to the phase transition for non-trivial $J(\eta)$.

In \cite{Maeda:2009wv}, it was shown that holographic superconductors have a universal value $\gamma =1$  for the critical exponent  associated with the thermodynamic susceptibility, $\chi_T\propto (T_c-T)^{-\gamma}$. The same analysis leads to $\gamma =1 $ also in the present more general models.
Putting these values together, we find that the critical exponents of the general Model II verify 
the Rushbrooke identity:
\be
\alpha +2\beta+\gamma =2\ .
\ee
Since $2<a_0\leq 4$, we find that $\beta\geq 1/2$
and $\alpha \leq 0$.
For $\alpha < 0$ (or  $2<a_0< 4$) the specific heat is continuous across the transition. 
This means that the transition is at least of third order (using the Ehrenfest convention that the order of the phase transition is the lowest one at which the derivative of the free energy is discontinuous). 
More generally, the phase transition is of order $n$ for
\be
n-1\geq  2\beta > n-2\ .
\ee
In other words, for these models the order of the phase transition is given by
\be
n=\lceil 2\beta + 1 \rceil =\lceil {a_0\over a_0-2}\rceil\ ,
\ee
where $\lceil  x \rceil $ denotes the smallest integer  greater or equal than $x$.
The transition will be of higher order than 3 if $a_0<3$. However, in this case, $\beta >1$ and the critical curve becomes
concave near $T_c$. 
These cases are of some theoretical interest because
there is a temperature $T_1<T_c$ where $d^2\langle O_i\rangle/dT^2$ changes sign. 
The change in the sign of  $d^2\langle O_i\rangle/dT^2$ could be an indicator of another phase transition, though we will not
further inquire on this point in this work.

It should be mentioned that {\it there are examples of real superconductors with third order phase transitions}
 (see e.g. \cite{Werner}). We are not aware of any real
superconducting system with well established transitions greater than third order (the experimental data given in
 \cite{Werner} are also consistent with the assumption that the transition is  fourth (or even higher) order; distinguishing the two cases
would require high experimental precision data which to our knowledge is not currently available). In our context, third order phase transitions
can be described by simple  models of type II (\ref{stmm}) with the addition of $\eta^3$ interactions in $G(\eta)$, $U(\eta)$ or $J(\eta)$.

For $a_0=4$, we have $\beta =1/2$, $\alpha=0$ and there is indeed a jump in the specific heat, which can be compared with the BCS result. For concreteness, let us focus on the case in which only $J=q^2\, \eta^2+j_0\,\eta^4$ is not trivial. Computing the jump in the specific heat, we obtain

\begin{equation}
\Delta\,c_v(T_c)=c_s(T_c)-c_n(T_c)=\frac{a^{(N)}_1\, j_0^2\, q^4+a^{(N)}_2\, j_0\, q^8+a^{(N)}_3\, q^{12}}{(q^4-a^{(N)}_4\, j_0)^3} \, ,
\end{equation}
where the $a^{(N)}_i$ are numerical factors whose values depend on the order $N$ to which we approximate the solution. The functional dependence of the specific heat on $q$ and $j_0$ does not depend on $N$. In particular, at $j_0=0$, i.e. for the HHH case, it is a constant independent of $q$. Furthermore, consistently with the analysis in section 3.1, 
we also find that $T_c\sim q$, in fact $T_c/q\approx 0.0587$ (see (\ref{mumu}) and below). Using numerical results to fit the proportionality coefficient, we obtain

\begin{equation}
\Delta c_v \approx \frac{9.5}{q}\, T_c \, .
\end{equation}
Interestingly, in BCS theory, the analogous quantity evaluates to 

\begin{equation}
\Delta c_v=\frac{8\, \pi^2}{7\, \zeta(3)}\, N(0)\, T_c\approx 9.38\, N(0)\, T_c \, ,
\end{equation}
where $N(0)$ is the density of states at the Fermi surface.
Comparing with (4.12) we see that, in this particular model, $1/q$ seems to play an analogous role as $N(0)$ (although there is of course no Fermi surface here). 

\section{Conductivity}

\label{section_conductivity}

So far we have been mainly concerned with the universality properties of the phase transitions described by our generalized models. As noted in subsection (2.2), these are controlled by the lowest order terms in the small condensate expansion of the functions $G(\eta)$, $U(\eta)$ and $J(\eta)$. Higher order terms can be then thought of as irrelevant perturbations of the conformal fixed point, thus not modifying the universality properties of the phase transition. However, they do play a key role in dynamical aspects away from $T_c$. Indeed, our models have a priori enough room to accommodate a plethora of behaviors that might model real systems. One of the most natural such dynamical quantities to consider is the conductivity. 
By the Kubo formula, it is related to the spectral density, which in turn represents the density of energy eigenstates at energy $\w $, weighted by the overlap with the
electric current operators.

Let us first briefly review how to study the conductivity in our general models. Following \cite{Gubser:2008px,Hartnoll:2008vx,Hartnoll:2008kx}, we consider time-dependent perturbations of $A_x= a_x(r)\ e^{-i\omega t}$ and $g_{tx}=f(r) \ e^{-i\omega t}$. These fluctuations are governed by the following equations
\bea
&& a''_x+ \left({g'\over g}-{\chi'\over 2}+{\partial_\eta G \, \eta'\over G}\right) a_x' +
\left({\omega^2\over g^2} e^\chi - {J\over gG}\right) a_x = {\phi'\over g}\ e^\chi \Big(-f'+{2\over r}\ f\Big)\ ,
\\
&& f' -{2\over r}\ f+G\ \phi' a_x =0\ .
\eea
 Substituting the second into the first equation, we find
 \be
a''_x+ \left({g'\over g}-{\chi'\over 2}+{\partial_\eta G \, \eta'\over G}\right) a_x' +\left( \Big({\omega^2\over g^2} -{G {\phi'}^2 \over g}\Big) e^\chi - {J\over gG}\right)a_x = 0\ .
\label{srl}
\ee
The asymptotic behavior of the perturbations is found to be
\be
a_x = a_x^{(0)}+ {a^{(1)}_x\over r} + \ldots \ ,\qquad f=r^2 f^{(0)} + {f^{(1)}\over r}+\ldots
\ee
 The conductivity can then be obtained from the formula
\be
\sigma = {J_x\over E_x} = - {ia_x^{(1)}\over \omega a_x^{(0)} }\ ,
\label{kond}
\ee
where in the second equality we have used the AdS/CFT dictionary.

\begin{figure}[h!]
\centering
\includegraphics[scale=0.62]{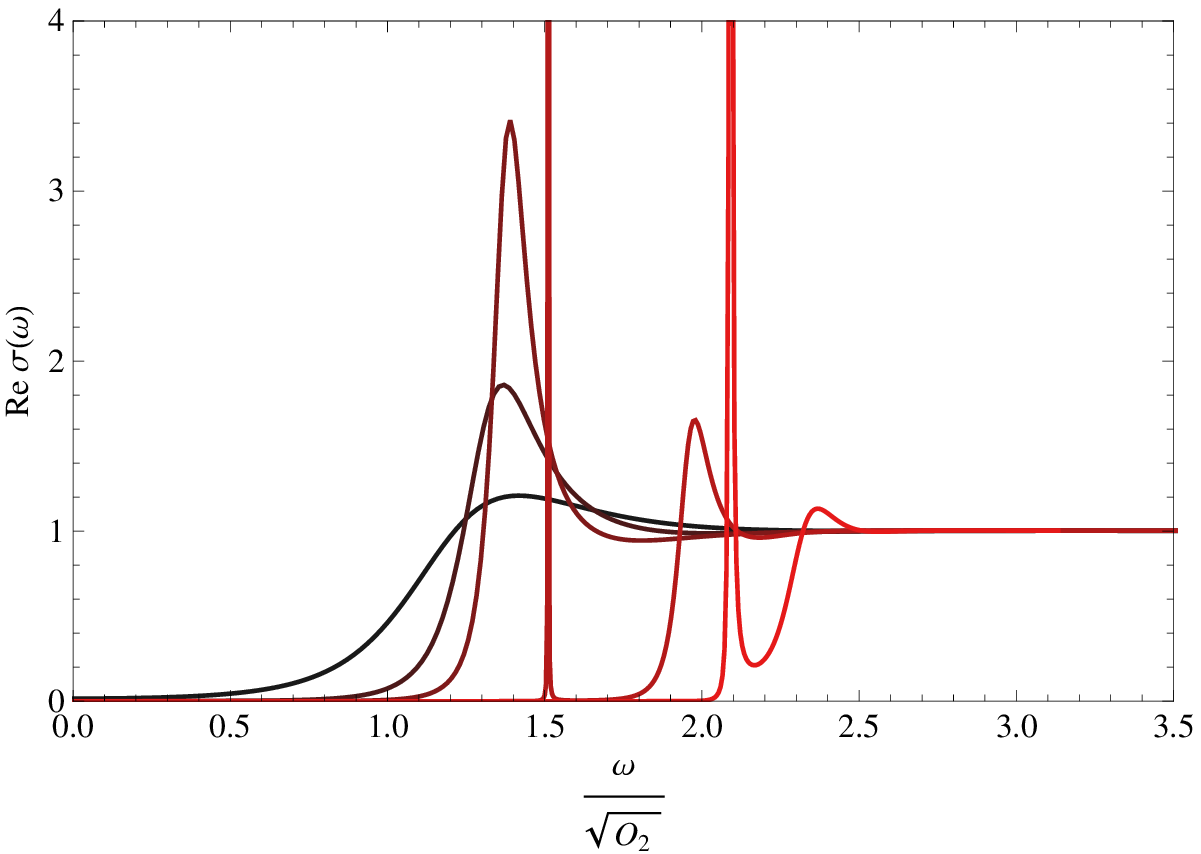} \hspace{.2cm}
\includegraphics[scale=0.62]{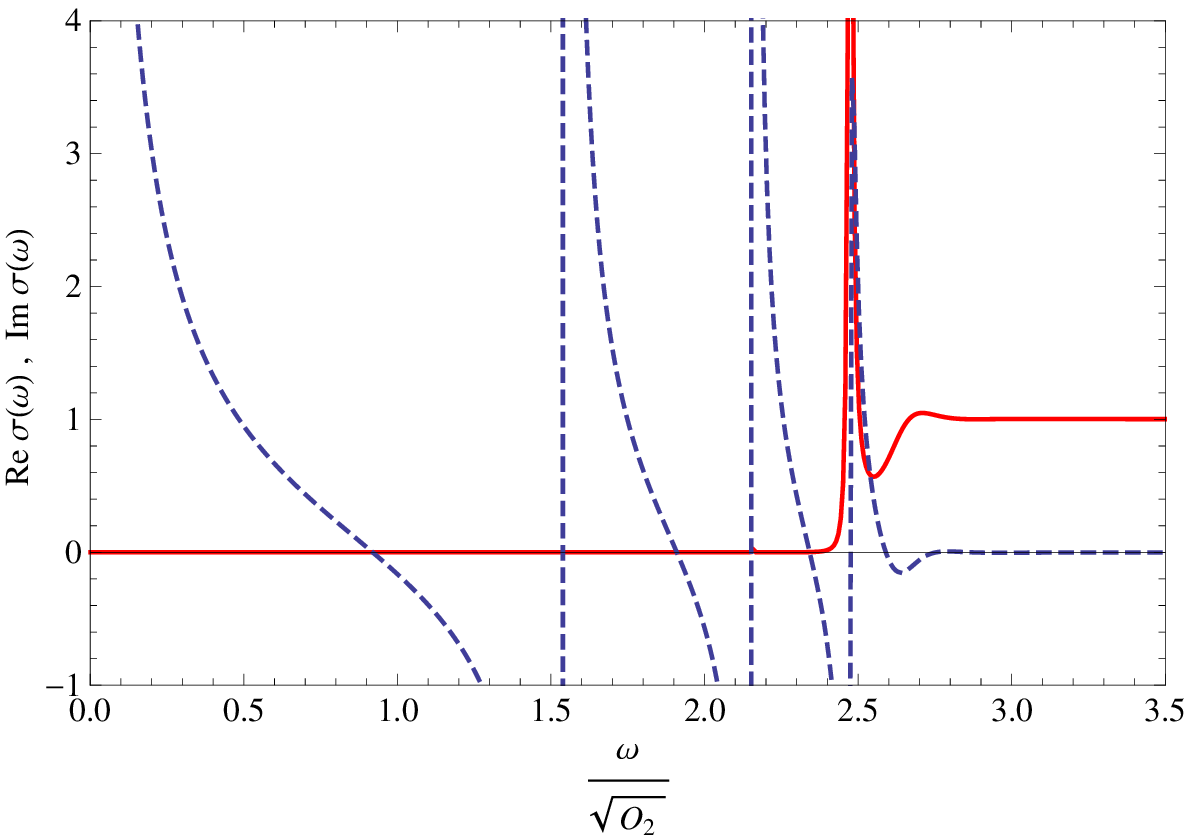} 
\caption{(a) Real part of the conductivity as a function of frequency for $J=\eta^2 +j_0 \eta^4$ with $j_0=0.6$  in the $\langle O_1\rangle = 0$ scheme. The curves correspond to different values of $T/T_c$ equal to $0.24,\ 0.29,\ 0.50,\ 0.61,\ 0.81$ (the curves with lower temperatures are those that go to zero more rapidly as $\w\to 0$). 
(b) Real and imaginary part of the conductivity for the same model at $T=0.20$.}
\label{conductivity_omega_and_T_curve}
\end{figure}

Following the ideas in section \ref{section_deformations}, we now investigate in detail the behavior of the conductivity under deformations that preserve dimension and charge of the condensate and the critical exponents. A simple implementation of this idea is given by the choice:
\beq
G=1\ ,\qquad U=1+{\eta^2\over 6}\ ,\qquad J=\eta^2 +j_0 \eta^ 4 \, ,
\label{J_conductivity}
\eeq
with $j_0\geq 0$. This choice provides a one-parameter family of deformations where $J$ remains positive for all $\eta$. For any given temperature, we can get arbitrarily close to the HHH model by reducing $j_0$.
Figure \ref{conductivity_omega_and_T_curve}  shows the conductivity for $j_0=0.6$ in the $\langle O_1\rangle = 0$ scheme (in this section we work at fixed charge density $\rho $).
The most salient feature of this plot is the appearance of resonance peaks in the conductivity, which increase in number and become narrower and higher as the temperature is gradually lowered. 
A similar effect is also seen at fixed temperature by increasing $j_0$.
One can also see that increasing $j_0$ leads to an increasing of the height of the peaks. The lowest temperature curve in fig. 3(a) is at $T=0.24$. Here the first peak in ${\rm Re }(\sigma)$ cannot
be seen numerically because it has become  narrower than the numerical grid. However, its presence can be inferred from the $1/(\w-\w')$ behavior of
${\rm Im}(\sigma)$, shown in fig. 3(b), through the Kramers-Kronig relation.

\begin{figure}[h!]
\centering
\includegraphics[scale=0.52]{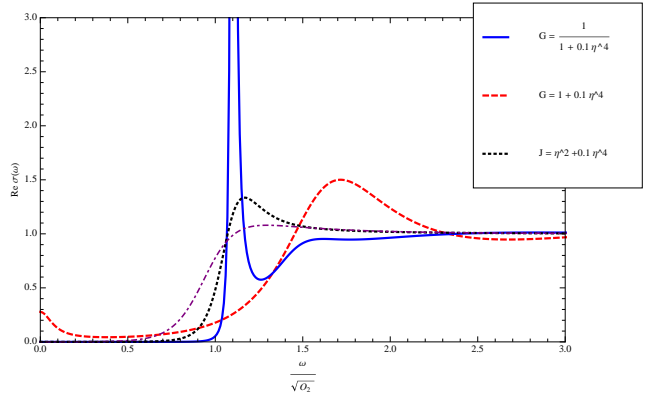} 
\caption{Conductivity as a function of frequency for HHH (dashed-dotted line) and for the different deformations given by the 
models (\ref{J_conductivity}), (\ref{JKKK}) and (\ref{JLLL}) in the $\langle O_1\rangle = 0$ scheme at $T=0.415\  T_c$.}
\label{comparaGJ}
\end{figure}

Consider now two different deformations:
\beq
G=1+g_0\eta^4\ ,\qquad U=1+{\eta^2\over 6}\ ,\qquad J=\eta^2 \, ,
\label{JKKK}
\eeq
or
\beq
G=\frac{1}{1+g_0\eta^4}\ ,\qquad U=1+{\eta^2\over 6}\ ,\qquad J=\eta^2 \, ,
\label{JLLL}
\eeq
with $g_0\geq 0$. 
The three models above belong to the same universality class, since they have
 the same critical temperature and the same critical exponents as in  the HHH model (obtained by setting $j_0=g_0=0$).
A natural question is to what extent transport properties are sensitive to the choice of deformations. 
Figure \ref{comparaGJ} compares the conductivities of the three different deformations (\ref{J_conductivity}), (\ref{JKKK}) and (\ref{JLLL}) in the $\langle O_1\rangle = 0$ scheme at a given temperature $T=0.415\ T_c$.
One can see that the conductivity undergoes significant changes relative to the HHH model, with strong dependence on the specific deformation. At this temperature, the model  (\ref{JLLL})   exhibits sharp peaks while the models (\ref{J_conductivity}), 
(\ref{JKKK}) have a smoother behavior, closer to BCS.
 In particular, the figure shows that
 a small deformation of the model can change the density of energy eigenstates of the system in a dramatic way.

Similar resonance peaks have been observed in \cite{Horowitz:2008bn}  by varying the dimension of the condensate 
(i.e. the mass of the scalar field), making it approach the BF bound. Our results show that it is not necessary to set the mass 
of the scalar field to any particular value. The same phenomenon can be obtained by higher order modifications of the functions $G(\eta)$, $U(\eta)$ and $J(\eta)$. 
In fact, this is not surprising: the conductivity peaks are a low temperature effect. As such, they are controlled by the value of the functions for 
large values of the condensate in contrast to e.g. critical exponents, which are determined by the small 
condensate expansion. In the simple models of \cite{Horowitz:2008bn}, which fit into
our class I of models, tuning the mass of the scalar $m$ (a rather drastic modification of the dual CFT) is the only available way of controlling the behavior of $J$ for any value of the condensate.

{}From the bulk perspective, the mechanism that gives origin to this peaks was clarified  in \cite{Horowitz:2009ij}. We now generalize this argument 
to our context. Introducing a new coordinate
\be
du = {e^{\chi/2}\over g}\ dr,
\ee
and defining $\Psi= \sqrt{G}\ a_x$,  (\ref{srl}) takes the form of a Schr\" odinger equation
\be
-{d^2\Psi\over du^2}+V(u) \Psi = \w^2 \Psi\ ,
\ee
with 
\be
 V\equiv g\big( G\, {\phi' }^2+ {J\over G}\ e^{-\chi}\big) + g^2e^{-\chi} \left({G''\over 2G}-{ {G'}^2\over 4G^2}+{G'\over 2G} \big({g'\over g} -{\chi'\over 2}\big)\right) \ ,
\label{potere}
\ee
where prime denotes derivative with respect to $r$.
It should be noted that $V$ not only depends on the couplings $G,\ J$, but also implicitly depends on $U$, through the dependence on $\eta ,\ \phi$ (which are in turn determined  by the coupled system
(\ref{Eq3uno}), (\ref{Eq4uno})). Note also that in the present approximation, where back reaction is neglected, $\chi $ can be set to zero and the first term proportional to ${\phi'}^ 2$ can be dropped
from (\ref{potere}).

\begin{figure}[h!]
\centering
\includegraphics[scale=0.65]{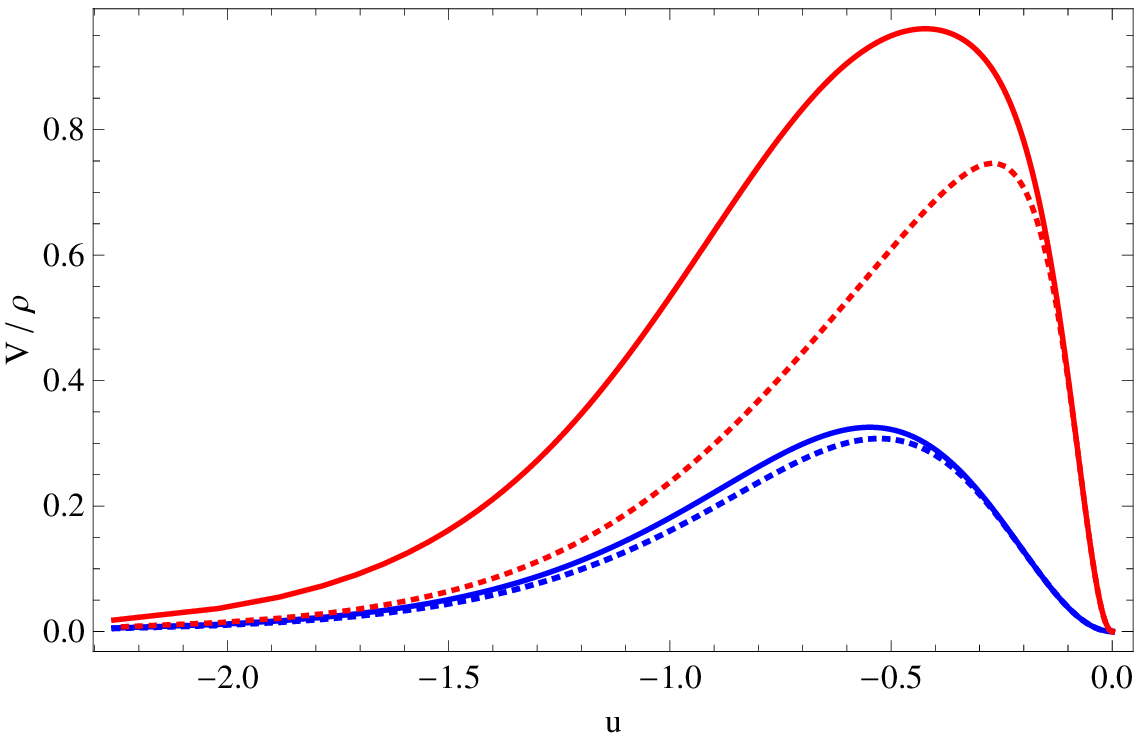} \hspace{.2cm} 
\includegraphics[scale=0.65]{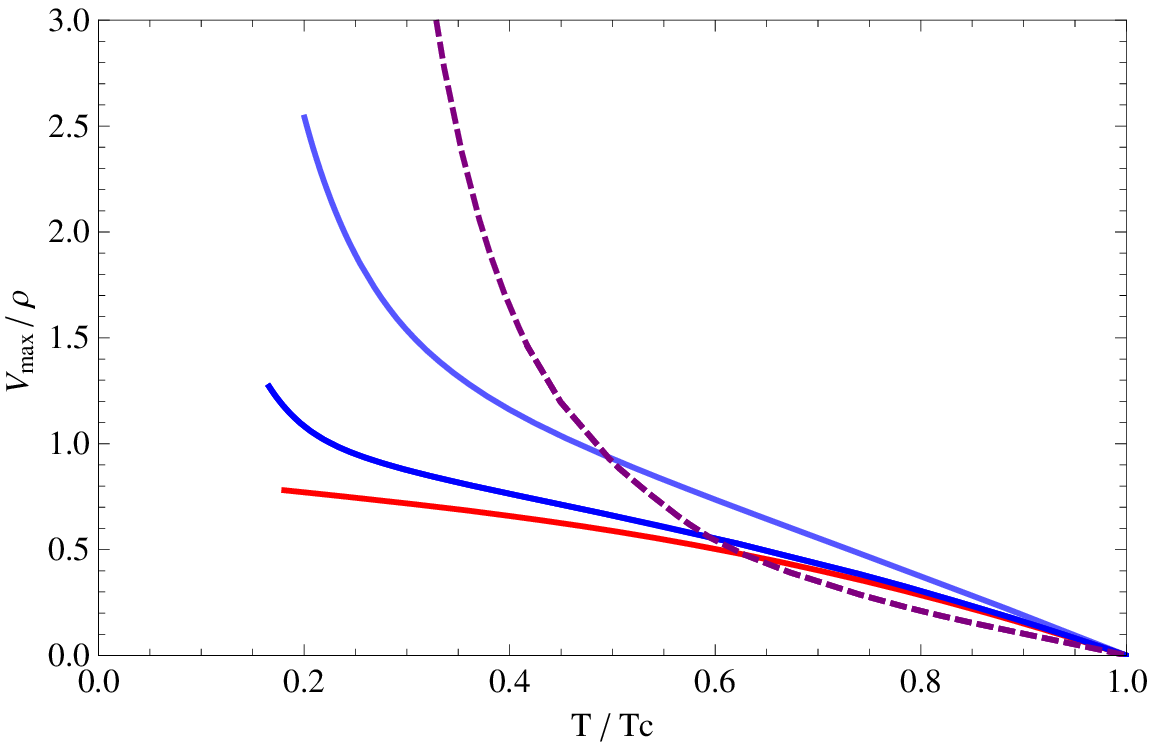} 
\caption{ (a) Potential for the model (\ref{J_conductivity}) with $j_0=0$  (dashed line), corresponding to the HHH model,  and for $j_0=0.05$ (solid line) in the scheme where 
$\langle O_2\rangle \neq 0$. 
The temperatures are  $0.78,\ 0.24$, from bottom to top. (b)
Maximum of the potential as a function of the temperature. The dashed line is the model (\ref{JLLL}) with $G=1/(1+0.1\eta^4)$; the solid lines correspond to (from bottom to top)
HHH model,  $J=\eta^2+0.05\eta^4$ and $J=\eta^2+0.2\eta^4$. }
\label{pote}
\end{figure}

Now consider for example the model (\ref{J_conductivity}).
The potential $V$ is shown in figure \ref{pote}(a), for $j_0=0.05$, along with the HHH case, $j_0=0$, always in the scheme where $\langle O_2\rangle \neq 0$.

The conductivity is obtained by sending an incoming wave from the right, which will be partly transmitted and partly reflected, with a reflection coefficient ${\cal R}$. 
As observed in \cite{Horowitz:2009ij} the conductivity (\ref{kond}) is nothing but
\be
\sigma (\w )= {1-{\cal R}\over 1+{\cal R}}\ .
\ee
If the potential is very high compared to  $\w^2$ (and as long as the area under the potential is also large), the absolute value of the reflection coefficient will be close to $1$.
The peaks are then produced when the reflected wave has a relative phase equal to $(2n+1)\pi$, so that $1+{\cal R}$ is near zero.
Figure \ref{pote} (b) shows the maximum of the potential $V_{\rm max}$ as a function of the temperature for different models. 
We see that $V_{\rm max}$ increases as the temperature is decreased.
 This gives room to new  resonance frequencies satisfying the condition $\w \ll \sqrt{V_{\rm max}}$ and hence the  emergence of new peaks. 
In HHH, in the scheme where $ \langle O_2\rangle$ is non-zero, there is a small region where the incoming wave is oscillating near $u=0$ and as a result there is a small
enhancement in ${\rm Re}(\sigma )$ at a certain frequency. Furthermore, $V_{\rm max}$ tends to a finite value at $T=0$ which is not big enough to  
allow for a reflected wave with $|{\cal R}|\sim 1$ and thus for the formation of peaks.

It is also worth noting that, as long as the potential stays finite at all temperatures (including $T=0$), there will be no hard gap in these models, since the probability of
tunneling will never be zero. As a result $|{\cal R}|<1$ and the conductivity will have a finite (albeit exponentially small) value at low temperatures.
However,  figure \ref{pote} (b) suggests that for some models $V_{\rm max}$ might actually go to infinity as $T\to 0$. If this is the case, there could be a hard gap and an infinite number of resonance peaks in this limit. The figure  \ref{pote} (b) also explains why  the model  $G=1/(1+0.1\eta^4)$  has a more suppressed conductivity at low frequencies and narrower peaks at a given temperature,  see fig. \ref{comparaGJ}; as
we can see, in this case, at a given temperature   the potential  barrier is higher.
We will make further comments on the issue of peaks in the Discussion section.

\section{Models with $\p_\eta \Theta\neq 0$ and  Hall effect without magnetic field}

\label{section_Hall}

We shall now incorporate the term (\ref{tet}). Even though a priori it contributes to the equations of motion as soon as $\p_\eta \Theta\neq 0$, it is easy to see that  for the particular ansatz (\ref{solu}) it actually gives a vanishing contribution.
Therefore we find exactly the same uncondensed and condensed black hole solutions irrespective of the coupling $\Theta (\eta )$.\footnote{The dyonic solution is also unaffected by the $\Theta $ term in the uncondensed phase, but it changes in the condensed phase if $\partial_\eta \Theta \neq 0$.} 
Nonetheless, this coupling will affect the conductivities in an important way.
The equation of motion for $A_y$ shows that this new interaction leads to the interesting effect that
an electric field in the $x$ direction turns on an electric field in the direction $y$, in very much the same fashion as in the Hall effect, this time without the presence of a magnetic field.
The basic idea behind this effect was first raised in \cite{hartnoll} in the context of a model having a constant $\Theta $. In that case the Hall conductivity is a numerical constant, present even at zero temperature and zero charge density.
The advantage of a coupling $\Theta (\eta)$ with $\p_\eta \Theta\neq 0$  is that it can incorporate non-trivial temperature and frequency dependence on the Hall conductivity $\sigma_{xy}$ 
through the non-trivial coupling to the scalar field. Moreover, we have the option of choosing $\Theta $ such that $\Theta(0 )=0$; in this case the $\Theta $ term
will be turned on only in the condensed phase. 
More recently, in \cite{Gauntlett:2009bh}, a non-trivial Hall effect was found by essentially the same mechanism as in this paper,  arising from a term of the form $f(h)\ F\wedge F$, where $h$ is an additional
real scalar field and $f$ is a specific function emerging from the compactification. 
A different way to obtain Hall conductivity appeared in the context of $p$-wave superconductors \cite{Roberts:2008ns}.

\subsection{Hall effect in the general model}

Let us consider the most general model with non-trivial $G(\eta)$,  $U(\eta)$ and $J(\eta)$. The relevant part of the action containing the $U(1)$ gauge field:
\be
S= \int d^{4}x\ \left(\sqrt{-\hat g}\left( -{1\over 4} G(\eta)\ F^{\mu\nu} F_{\mu\nu}-{1\over 2} J(\eta) A_{\mu} A^{\mu}\right)+\ 
\frac{1}{4}\Theta(\eta) \ \epsilon^{\mu\nu\rho\sigma}F_{\mu\nu}F_{\rho\sigma}\right)\ .
\label{wzero}
\ee
The variation with respect to $A_{\sigma}$ leads to a generalized version of the London  equation
\be
\p_{\mu}\sqrt{-\hat g}\ G(\eta)F^{\mu\sigma}-\sqrt{-\hat g}\ J(\eta)\ A^{\sigma}-\epsilon^{\mu\nu\rho\sigma }F_{\mu\nu}\p_{\rho}\Theta(\eta)=0\ .
\label{wuno}
\ee
Using the ansatz (\ref{solu}) we find that the equation (\ref{Eq3}) for the potential $\phi $ remains unchanged.
In order to calculate the conductivity we consider a perturbation $A_\mu=A_\mu(t,r)$ to the background. 
We will ignore backreaction and consider as usual the linear order. The effect produced by the new $\Theta$ term in (\ref{wuno}) is to couple (at first order) the bulk perturbations $A_x$ and $A_y$. Indeed the equations of motion are ($\epsilon^{t r x y}=1$)
\be
\p_r^2A_x+\Big(\ \frac{\p_r g}{g}+\frac{\p_rG}{G}\ \Big)\p_rA_x
-\frac{1}{g^2}\p_t^2A_x-\frac{J}{gG}\ A_x-\frac{2}{gG(\eta)}\p_tA_y\p_r\Theta\ =0\ ,
\ee
\be
\p_r^2A_y+\Big(\ \frac{\p_r g}{g}+\frac{\p_rG}{G}\ \Big)\p_rA_y
-\frac{1}{g^2}\p_t^2A_y-\frac{J}{gG}\ A_y+\frac{2}{gG(\eta)}\p_tA_x\p_r\Theta\ =0\ .
\ee
Expanding in Fourier modes
\be
A_x=\int d\w \ e^{-i\w t} a_x(r; \w)\ , \qquad  A_y =\int d\w \ e^{-i\w t} a_y (r; \w)\ , 
\label{wsei0}
\ee
the equations for the modes $a_x(r; \w), \ a_y(r; \w)$ become
\be
a''_x+\Big(\ \frac{g'}{g}+\frac{G'}{G}\ \Big) a'_x+
\Big(\frac{\w^2}{g^2}-\frac{J}{gG}\Big)a_x+2i \w \frac{\Theta '}{gG(\eta)}\ a_y =0\ ,
\label{wsei1}
\ee
\be
a''_y+\Big(\ \frac{g'}{g}+\frac{G'}{G}\ \Big) a'_y+
\Big(\frac{\w^2}{g^2}-\frac{J}{gG}\Big)a_y-2i\w \frac{\Theta '}{gG(\eta)}\ a_x =0\ ,
\label{wsei2}
\ee
where prime indicates differentiation with respect to $r$ (so that $G'=\p_\eta G\ \eta',\ \Theta' = \p_\eta \Theta \ \eta'$).
The system (\ref{wsei1})-(\ref{wsei2}) can be decoupled introducing complex coordinates, $z=x+iy$, so that 
$$
A_z ={1\over 2}( a_x- ia_y)\ ,
\qquad A_{\bar z} ={1\over 2}( a_x+ ia_y)\ .
$$ 
Using the notation,
\be
\hat{P}=\p_r^2+\Big(\frac{g'}{g}+\frac{G'}{G}\Big)\p_r+\left(\frac{\w^2}{g^2}-\frac{J}{gG}\right),\qquad
\hat{Q}=2\w \frac{\Theta '}{gG(\eta)}\ ,
\ee
the equations take the form
\be
\hat{P}A_{z}-\hat{Q}A_{z}=0\ ,\qquad \hat{P}A_{{\bar z}} +\hat{Q}A_{{\bar z}}=0\ .
\label{wsette}
\ee
Note that the asymptotic behavior of (\ref{wsette}) is not modified by the new term; indeed for $r\rightarrow\infty$ we have
\be
A''_{z}+\frac{2}{r}A'_{z}- 2\w \frac{\p_{\eta}\Theta}{r^2}\eta'\ A_{ z}=0\ , \qquad 
A''_{\bar z}+\frac{2}{r}A'_{\bar z}+ 2\w \frac{\p_{\eta}\Theta}{r^2}\eta'\ A_{\bar z}=0\ .
\ee
Since $\eta'$ is $O(1/r^2)$ or $O(1/r^3)$, the $\Theta $ term can be neglected at large $r$. 
Then the asymptotic solutions are the same as in the  $\Theta=0$ case,
\be
A_z=A_z^{(0)}+\frac{A_z^{(1)}}{r}\ ,\qquad A_{\bar z}=A_{\bar z}^{(0)}+\frac{A_{\bar z}^{(1)}}{r}\ .
\label{wsette1}
\ee
Finally we observe the fact that the equations (\ref{wsette}) do not depend on the value of $\Theta(0)$.

To compute causal behavior, we solve for the fluctuations with ingoing-wave boundary conditions at the horizon. This requires 
\be
A_z\sim\ C_z\ \Big(1-\frac{r}{r_h}\Big)^{-i\w/3}a_z(r)\ ,\qquad A_{\bar z}\sim\ C_{\bar z}\ \Big(1-\frac{r}{r_h}\Big)^{-i\w/3}a_{\bar z}(r)\ ,
\ee
with $a_z(r)=1+a_z^{(1)}(1-r/r_h)+\ldots$ and similarly for $a_{\bar z}$.%
\footnote{Equations (\ref{wsette}) contain two different sources of singularities at the horizon, coming from inverse powers of $g^2\sim (1-r/r_h)^2$ and $g\sim (1-r/r_h)$. The exponent $-i\w$/3 takes care of the first one, while the second one is canceled by choosing appropriates values for $a_{z}^{(1)}$ and $a_{\bar z}^{(1)}$. This determines one integration constant in (\ref{wsette}),  whereas  $C_z$ and $C_{\bar z}$ remain as free parameters.} Because (\ref{wsette}) are linear and homogeneous equations we can first set $C_z=C_{\bar z}=1$ and find two solutions 
$ \A_{z}$, $ \A_{\bar z}$. Then the most general solutions are obtained as
\be
A_z=C_z \A_z ,\qquad A_{\bar z}=C_{\bar z} \A_{\bar z} \ .
\label{asies}
\ee
From the above expressions we find,
\be
a_x=C_{\bar z}\A_{\bar z}+C_z\A_z\ ,\qquad
a_y=-i \left(C_{\bar z}\A_{\bar z}-C_z\A_z \right)\ .
\ee
To uncover the physical meaning of $C_{z}$ and $C_{\bar z}$, we consider the asymptotic behavior (\ref{wsette1}) and find
\be
a_x (\w )= a_x^{(0)}+\frac{a_x^{(1)}}{r}\ ,\qquad a_{y}(\w)=a_{y}^{(0)}+\frac{a_{y}^{(1)}}{r}\ ,
\ee
\be
a_x^{(0)}=C_{\bar z}\A_{\bar z}^{(0)}+C_z\A_z^{(0)}\ ,\qquad a_x^{(1)}=
C_{\bar z}\A_{\bar z}^{(1)}+C_z\A_z^{(1)}\ ,
\ee
\be
a_y^{(0)}=-i \big( C_{\bar z}\A_{\bar z}^{(0)}-C_z\A_z^{(0)}\big)\ ,\qquad a_y^{(1)}=
-i\big(C_{\bar z}\A_{\bar z}^{(1)}-C_z\A_z^{(1)}\big)\ .
\label{wotto}
\ee
As is standard in AdS/CFT, the leading term in the asymptotic expansion of the fields is related to the source in the dual theory; therefore, from the definition $E_i=-\p_t A_i$, $i=x,y$, we obtain the system,
\be
\left\{\begin{array}{l}
E_x\ =\ i\w a_x^{(0)}=\ C_{\bar z}i\w \A_{\bar z}^{(0)}+C_z i\w\A_z^{(0)}\\\\
iE_y=i^2\w a_y^{(0)}=C_{\bar z}i\w \A_{\bar z}^{(0)}-C_zi \w \A_z^{(0)}
\end{array}\right. 
\ee
In this way, the integration constants $C_{z}$ and $C_{\bar z}$ get related to the physical sources $E_x$ and $E_y$. Solving for $C_{z}$ and $C_{\bar z}$ we find the expressions
\be
C_{\bar z}=\frac{E_x+iE_y}{2i\w\A_{\bar z}^{(0)}}\ ,\qquad 
C_{z}=\frac{E_x-iE_y}{2i\w\A_{z}^{(0)}}\ .
\ee
The asymptotic coefficients $a_x^{(1)}$ and $a_y^{(1)}$ in the expansion (\ref{wotto}) can now be written in terms of the electric field components
as
\be
a_x^{(1)}=
\frac{1}{i\w}\left(\frac{\A_{\bar z}^{(1)}}{\A_{\bar z}^{(0)}}+\frac{\A_z^{(1)}}{\A_z^{(0)}}\right)\frac{E_x}{2}+
\frac{1}{\w}\left(\frac{\A_{\bar z}^{(1)}}{\A_{\bar z}^{(0)}}-\frac{\A_z^{(1)}}{\A_z^{(0)}}\right)\frac{E_y}{2}\ ,
\label{wotto1}
\ee
\\
\be
\ a_y^{(1)}=
-\frac{1}{\w}\left(\frac{\A_{\bar z}^{(1)}}{\A_{\bar z}^{(0)}}-\frac{\A_z^{(1)}}{\A_z^{(0)}}\right)\frac{E_x}{2}+
\frac{1}{i\w}\left(\frac{\A_{\bar z}^{(1)}}{\A_{\bar z}^{(0)}}+\frac{\A_z^{(1)}}{\A_z^{(0)}}\right)\frac{E_y}{2}\ .
\label{wotto2}
\ee
This expression will be the starting point for the discussion about the conductivity.

\bigskip

It is convenient to rewrite the action (\ref{wzero}) in the form
\be
S= \int d^{4}x\ \left(\sqrt{-\hat g}\left( -{1\over 2} G(\eta)\ F^{\mu\nu} \p_{\mu}A_{\nu}-{1\over 2} J(\eta) A_{\mu} A^{\mu}\right)+\ 
\frac{1}{2}\Theta(\eta) \ \epsilon^{\mu\nu\rho\sigma}F_{\mu\nu}\p_{\rho}A_{\sigma}\right)\ .
\label{wnove}
\ee
{}From   (\ref{wuno}) we have the relation
\be
-\frac{1}{2}\sqrt{-\hat g}J(\eta)A^{\mu}A_{\mu}=-\frac{1}{2}\p_{\mu}\big(\sqrt{-\hat g}G(\eta)F^{\mu\nu}\big)A_{\nu}
+\frac{1}{2}\epsilon^{\mu\nu\rho\sigma}F_{\mu\nu}\p_{\rho}\Theta A_{\sigma}=0\ .
\label{wdieci}
\ee
Substituting (\ref{wdieci}) into (\ref{wnove}) we find

\be
S_{o.s.} = -\frac{1}{2}\int d^{4}x\ \Big[\ \p_{\mu}\Big(\sqrt{-\hat g}\ G(\eta)F^{\mu\nu}A_{\nu}\Big)-
\p_{\mu}\Big(\Theta(\eta)\epsilon^{\mu\nu\rho\sigma}F_{\nu\rho}A_{\sigma}\Big)\ \Big]\ .
\ee
where $S_{o.s}$ denotes the on-shell action.
We now specify our ansatz. We obtain 
\bea
S_{o.s.}\ &=&\ -\frac{1}{2}\int d^{3}x\ \Big[\ \sqrt{-\hat g}\ G(\eta)F^{r\nu}A_{\nu}-
\Theta(\eta)\epsilon^{r\nu\rho\sigma}F_{\nu\rho}A_{\sigma}\ \Big]_{r=r_h}^{r\rightarrow\infty} \nonumber\\ 
\nonumber\\ 
&=&-\frac{1}{2}\int d^{3}x\ \Big[\ g(r)\big(A_x\p_rA_x+A_y\p_rA_y\big)+\Theta(\eta)\big(A_y\p_tA_x-A_x\p_tA_y\big)\ \Big]_{r=r_h}^{r\rightarrow\infty}
\eea
Using the boundary condition $A_{x}(r_h)=A_{y}(r_h)=0$ and the Fourier representation (\ref{wsei0}) we find
\bea
S_{o.s.} &=& -\frac{1}{2}
\int d^2x\int d\w\ \Big[\ g(r)\big(a_x(r,\w)\p_r a_x(r,-\w)+a_y(r,\w)\p_r a_y(r,-\w)\big)\Big]^{r\rightarrow\infty}
\non\\
&-& \frac{1}{2}\int d^2x\int d\w\ \Theta(\eta)\Big[\ i\w a_y(r,\w)a_x(r,-\w)-i\w a_y(r,-\w)a_x(r,\w)\ \Big]^{r\rightarrow\infty}\ 
\eea
{}Finally, substituting the asymptotic behavior for the fields, we obtain the result,
\be
S_{o.s.} = \frac{1}{2}
\int d^2x\int d\w\ \Big[a_x^{(0)}(\w ) a_x^{(1)}(-\w)+a_y^{(0)}(\w)a_y^{(1)}(-\w)+i\w\Theta(0)\epsilon^{ij}a^{(0)}_i(\w)a^{(0)}_j(-\w)\ \Big]\ ,
\label{Sonshell}
\ee
where $i,j=x,y$ and we use the convention $\epsilon^{xy}=1$.

In the AdS/CFT dictionary, the leading term in the asymptotic expansion of the fields determines a source in the dual theory, while the ``normalizable'' term will give the expectation value of the dual current. Therefore, from (\ref{wotto}) and the definition $E_i=-\p_t A_i$, we get,
\bea
J_x&=&\frac{\delta S_{o.s.}}{\delta a_x^{(0)}}= 
a_x^{(1)}(\w )-i\w\Theta(0)a^{(0)}_y(\w)=a_x^{(1)}(\w )-\Theta(0)E_y\ , \label{w11}\\
\nonumber\\
J_y&=&\frac{\delta S_{o.s.}}{\delta a_y^{(0)}}= 
a_y^{(1)}(\w )+i\w\Theta(0)a^{(0)}_x(\w)=a_y^{(1)}(\w )+\Theta(0)E_x\ . 
\label{w12}
\eea
Using the expressions (\ref{wotto1})-(\ref{wotto2}) we derive the components of the conductivity matrix,
defined by $J_i =\sigma_{ij}E_j$,
\bea
&& \sigma_{xx}=\frac{1}{2}\left(\sigma_{zz}+ \sigma_{\bar z\bar z}\right)
=\sigma_{yy}\ ,
\non\\
&& \sigma_{xy}= \frac{1}{2i}\left(\sigma_{zz} - \sigma_{\bar z\bar z}\right) = 
-\sigma_{yx}\ ,
\label{sarah}
\eea
where $\sigma_{zz}$,  $\sigma_{\bar z\bar z} $ represent the conductivities for left-oriented and right-oriented circular polarizations of the electric field,
\be
\sigma_{zz}= \frac{1}{i\w }\frac{\A_z^{(1)}}{\A_z^{(0)}}\ ,\qquad 
\sigma_{\bar z\bar z}= \frac{1}{i\w }\frac{\A_{\bar z}^{(1)}}{\A_{\bar z}^{(0)}}\ ,\qquad \sigma_{z\bar z}=0\ .
\label{arans}
\ee
The relations $\sigma_{xx}=\sigma_{yy}$ and $\sigma_{xy}=-\sigma_{yx}$ are a consequence of isotropy.
In a parity-preserving theory,
$\sigma_{zz}$ and  $\sigma_{\bar z\bar z} $ are equal to each other, and the Hall conductivity $\sigma_{xy}$ (proportional to the difference) vanishes.
On the contrary, in a parity-violating theory, 
they differ, giving rise to a non-trivial Hall effect:
turning on an external electric field in the $x$ direction implies that the system must automatically produce an electric field in the $y$ direction
and a non-trivial $J_y$ current (and viceversa).
{}For a generic $\Theta (\eta )$, parity symmetry is explicitly broken in the  present model and we  indeed obtain a non-trivial Hall conductivity $\sigma_{xy}$,
even if $\Theta (0)=0$. In turn, as described above, for special couplings of the form $\Theta = \theta_0 \eta^{2k+1}$, with integer $k$, parity and time-reversal symmetries still hold if we assume that
$\eta $ is a pseudoscalar transforming as $\eta\to -\eta $ under $P$ and $T$.
However, in this case, in the condensed phase the non-trivial profile of $\eta $ spontaneously breaks the parity and time-reversal symmetries, thus leading again to non-trivial Hall conductivity.

In the uncondensed phase, $\eta\equiv 0$, and the parity-violating interaction is not turned on.
The system has vanishing Hall conductivity as expected.

\subsection{Numerical analysis of the conductivities}

Here we shall consider as an example the HHH action  \cite{Hartnoll:2008vx} with the addition of the term (\ref{tet}), with
 $\Theta =\eta ^n $. 
The total action reads
\bea
S_0 &=& {1\over 16\pi G_N}\int d^{3+1}x \bigg( \sqrt{-\hat g}\left( R -{1\over 4}  F^{\mu\nu} F_{\mu\nu} +{6\over L^2} (1+ {1\over 6} \eta^2 ) -{1\over 2} 
(\partial \eta)^2 -{1\over 2} q^2 \eta^2 (\partial_\mu \theta -A_\mu )^2\right)
\non\\
&+& {1\over 4}\ \eta^n  \ \epsilon^{\mu\nu\rho\sigma}F_{\mu\nu}F_{\rho\sigma}\bigg)\ .
\label{ccero}
\eea
We take $q=1$ and $L=1$ and, once more, fix the gauge $\theta =0$.
The $n=1$ case is of interest, being the simplest case, and also because  P and T symmetries are preserved. 
The  $n=2$ case is also of interest, since this case is easily 
incorporated in the context of HHH model with a complex scalar field $\psi$ by adding to the HHH model the interaction
$\psi^*\psi\ F\wedge F$.

\begin{figure}[tbh]
\centering
\subfigure[]{\includegraphics[width=7.5cm]{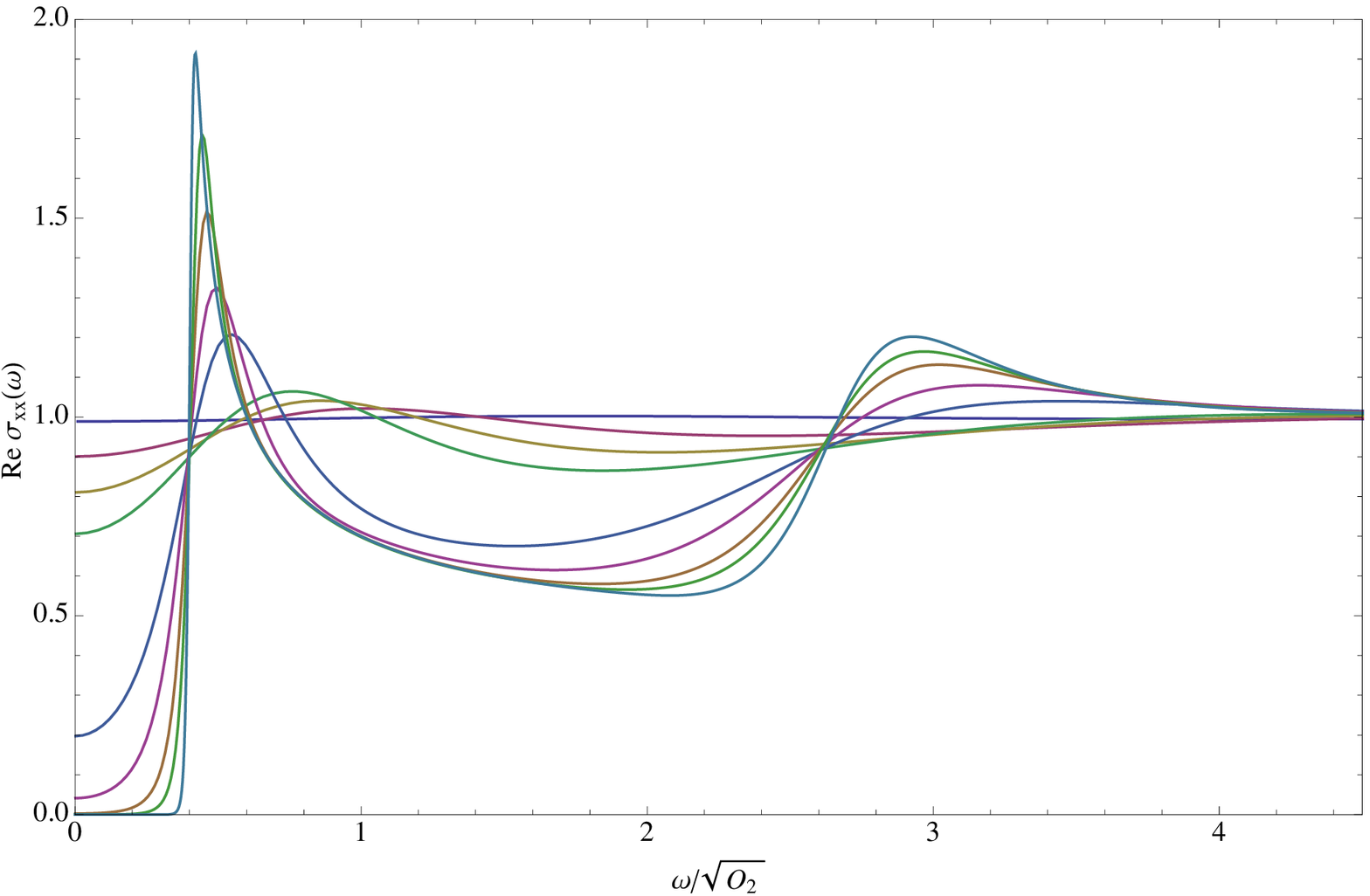}}
\subfigure[]{\includegraphics[width=7.5cm]{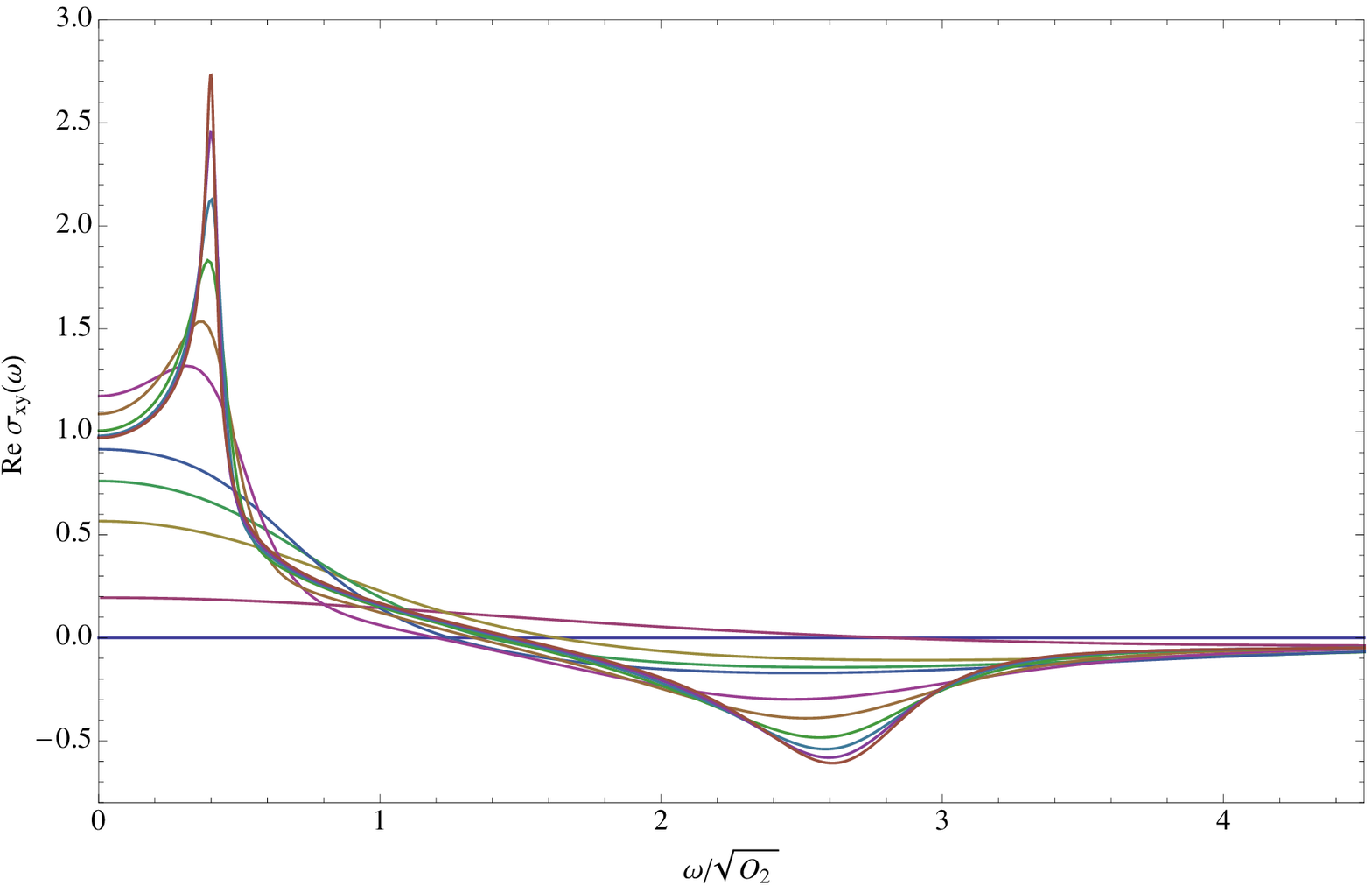}}
\caption{(a)  ${\rm Re}(\sigma_{xx})$ vs. $\w $ (lower temperatures corresponds to curves with lower intercepts at $\w =0$) and
(b) ${\rm Re}(\sigma_{xy})$ vs. $\w $ for the model (\ref{ccero}) with $n=1$  at various temperatures, in the scheme where $\langle O_2\rangle $ is non-zero
 (here the curves with lower temperatures are those which have higher value of  ${\rm Re}(\sigma_{xy})$ at the frequency of the peak).
\label{sgHall}}
\end{figure}

\begin{figure}[tbh]
\centering
\subfigure[]{\includegraphics[width=7.5cm]{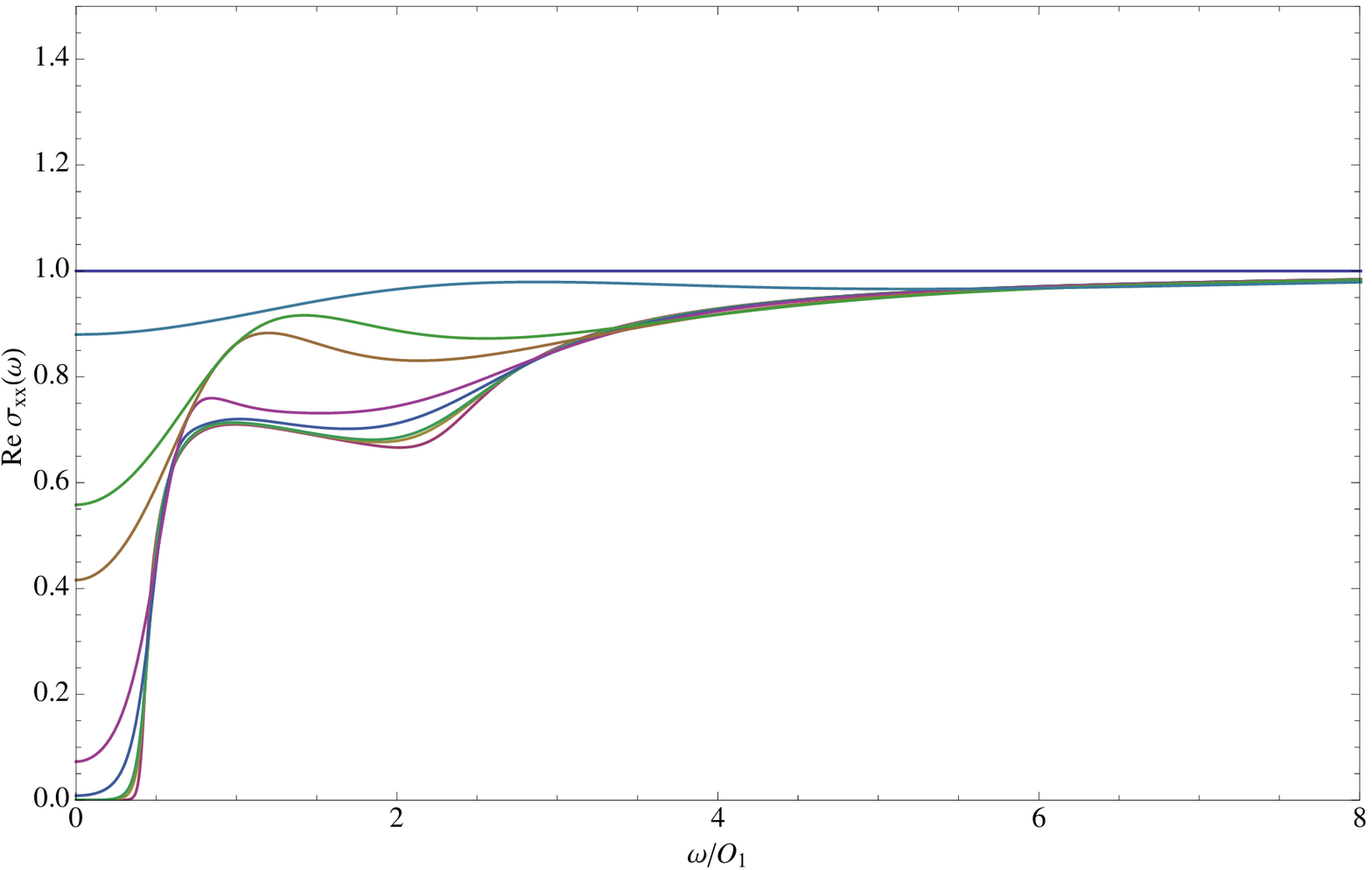}}
\subfigure[]{\includegraphics[width=7.5cm]{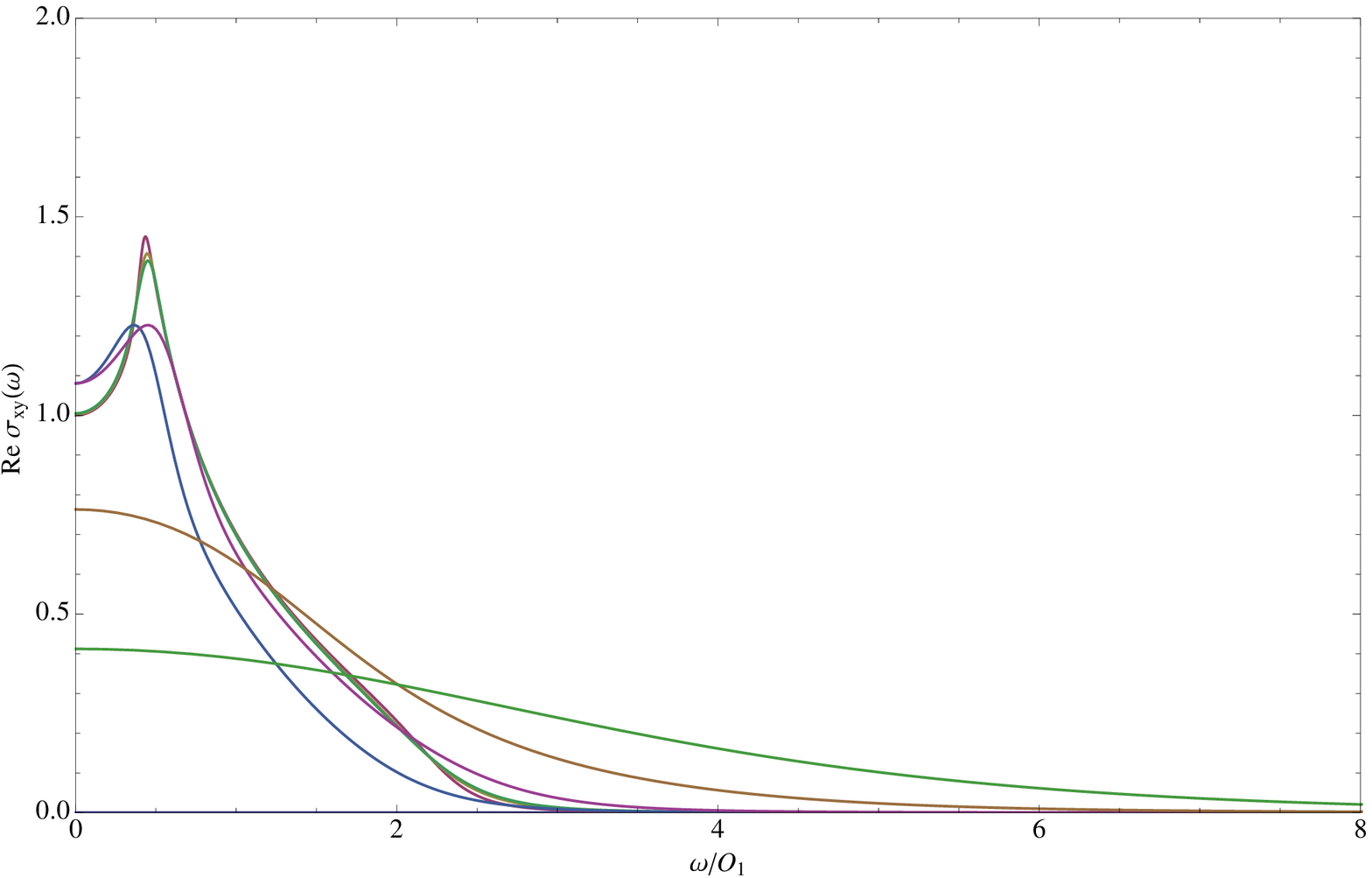}}
\caption{(a)  ${\rm Re}(\sigma_{xx})$ vs. $\w $ and
(b) ${\rm Re}(\sigma_{xy})$ vs. $\w $ for the model (\ref{ccero}) with $n=1$ at various temperatures, in the scheme where $\langle O_1\rangle $ is non-zero 
(same conventions as in figure \ref{sgHall}).
\label{sg1Hall}}
\end{figure}

Our aim is to compute $\sigma_{xx}$ and $\sigma_{xy}$ using (\ref{sarah}) and (\ref{arans}), by numerically solving the differential equations (\ref{wsette}),
which for this particular choice of couplings read
\bea
&& A''_z+\frac{g'}{g}A'_z+\left(\frac{\w^2}{g^2}-\frac{\eta^2}{g}-2n \w \eta^{n-1} \frac{\eta '}{g} \right)A_{z}=0\ ,
\non\\
&& A''_{\bar z}+\frac{g'}{g}A'_{\bar z}+\left(\frac{\w^2}{g^2}-\frac{\eta^2}{g}+2n\w \eta^{n-1}\frac{\eta '}{g} \right)A_{\bar z}=0\ . 
\label{roni}
\eea
with $g$ given in (\ref{geres}). Solving these equations numerically with the boundary conditions as 
described in the previous section, we compute the conductivities (\ref{sarah}) and (\ref{arans}) at different temperatures.
The results for the model with $n=1$ are shown in figures \ref{sgHall}(a),(b),  obtained in the scheme where $\langle O_1\rangle =0$, and
in figures \ref{sg1Hall}(a),(b),  obtained in the scheme where $\langle O_2\rangle =0$.
 At high frequencies, ${\rm Re}(\sigma_{xx})$ approach the normal phase behavior seen at $T=T_c$, showing that the
mechanism for conduction is through the normal phase charge carriers, while ${\rm Re}(\sigma_{xy})$ approaches zero in this high frequency regime.
As the temperature is lowered, there is a gap in the direct conductivity ${\rm Re}(\sigma_{xx})$  at $\w< \w_g$, with $\w_g \sim \sqrt{\langle O_2\rangle }$ and
$\w_g \sim \langle O_1\rangle$ in the two different schemes.
This is expected, $\w_g$ representing the minimal energy to break the Cooper pairs. On the other hand
 ${\rm Re}(\sigma_{xy})$ has no gap; it approaches a finite value as $\w\to 0$.
In the limit that the temperature goes to zero the curve   ${\rm Re}(\sigma_{xx})$ in  \ref{sgHall}(a) exhibits a pronounced resonance peak at a frequency $\w $ slightly higher than $\w_g$. 
We also see a mild minimum and a mild maximum at some higher values of the frequency.
In the other scheme where $\langle O_1\rangle \neq 0$, we find a smooth behavior after the gap.
The model with $n=2$ reproduces qualitatively similar features but with many more peaks.

We have also verified the following important feature. While the direct conductivity ${\rm Re}(\sigma_{xx})$ has the expected delta function singularity at $\w=0$, the Hall conductivity ${\rm Re}(\sigma_{xy})$ has a finite
value in the DC  ($\w=0$) case.
This is seen more clearly from the behavior of ${\rm Im}(\sigma_{xx})$ and  ${\rm Im}(\sigma_{xy})$ near $\w =0$: by the Kramers-Kronig relation, a pole in the imaginary
part of the conductivities implies a delta function singularity in the real part (see e.g. \cite{Hartnoll:2008vx}).
The numerical analysis shows that ${\rm Im}(\sigma_{xx})\sim 1/\w$ while  ${\rm Im}(\sigma_{xy})\sim \w^0$ as $\w\to 0$.

\begin{figure}[tbh]
\centering
{\includegraphics[width=7.5cm]{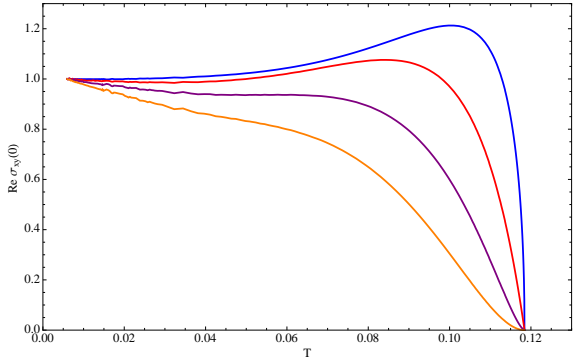}}
\caption{${\rm Re}(\sigma_{xy})(\w=0)$ vs. $T$ for the model (\ref{ccero}) with $n=1$, $n=2$, $n=3$ and $n=4$ (from top to bottom).
Since the different curves correspond to different theories, for comparison ${\rm Re}(\sigma_{xy})(\w=0)$ has been normalized to 1 at the lowest temperature numerically available.}
\label{THall}
\end{figure}

Having a finite DC Hall conductivity, it is of interest to investigate its dependence with the temperature.
This is shown in figure \ref{THall} for the $n=1,2,3,4$ models. The first obvious feature is that
the Hall conductivity vanishes at $T\geq T_c$,
 since in this case $\eta =0$ and the generalized ``theta term" does not contribute.
 Near $T_c$, in general the DC Hall conductivity has the behavior
 \be
 {\rm Re}(\sigma_{xy})(\w=0)\sim (T_c-T)^\nu\ ,
 \ee
where $\nu $ depends on $n$ (e.g. $\nu\sim 0.4$ and $\nu\sim 1$ for $n=1,2$ respectively, and $\nu>1$ for $n>2$).
In the $n=1,2,3$ cases, at lower temperatures the DC Hall conductivity is approximately temperature independent. 

\section{Discussion}

In the first part of this paper we have explored the properties of the holographic superconductor system
described by the action (\ref{cero}), for different couplings $G(\eta )$, $U(\eta )$ and $J(\eta )$
with the small condensate behavior (\ref{stm}) or (\ref{stmm}). We confirmed that models of the type I (\ref{stm}) have second order 
phase transition with mean field theory exponents, while models of type II can have phase transitions
of arbitrary order and critical exponents that depend on the parameters of the model. We analytically determined this dependence finding the following behavior
\be
\langle O\rangle \sim (T_c-T)^\beta\ ,\qquad \Delta c_v \sim (T_c-T)^{-\alpha }\ ,
\qquad \chi_T\sim (T_c-T)^{-\gamma}\ ,
\ee
\be
\beta ={1\over a_0-2}\ ,\qquad \alpha=-{4-a_0\over a_0-2}\ ,\qquad \gamma =1\ ,
\ee
with
\be
a_0={\rm Min}\{ a,b,c\}\ ,\qquad 2<a_0\leq 4\ .
\ee
It follows that the standard Rushbrooke identity among critical exponents, $\alpha +2\beta +\gamma =2 $,
is satisfied. This is a reassuring feature, since as we noted in the text, many of these models most probably do not admit a String/M theory embedding. The fact that the Rushbrooke identity is satisfied suggests that at least at large $N_c$ the  phenomenological models II are still describing a field theory-like behavior. 

The order of the phase transition is $n=\lceil {a_0\over a_0-2}\rceil$\ ,
with $\lceil  x \rceil $ denoting the smallest integer  greater or equal than $x$.
This permits to accommodate systems with critical exponents $\beta\geq 1/2 $, the  field value $\beta =1/2$ corresponding to interactions containing
even powers of $\eta $. Another particular case is $\beta =1$, which appears in  models  containing (analytic) interactions $\eta^3$ (i.e. $a_0=3$).
These models undergo third order phase transitions.
This type of behavior can also be incorporated in the Landau-Ginzburg framework in the absence of an $\eta \leftrightarrow -\eta$ symmetry.

We have also studied the conductivity in models obtained by deformations of the HHH model by terms $O(\eta^4)$.
 The conductivity reproduces
essentially the same basic behavior as in the HHH model, namely a gap at $\w<\w_g$ and at high frequencies it
approaches the plateau of the normal phase.
However, in some models the conductivity exhibits resonance peaks in frequency, which are similar to  those that have previously appeared in the 4+1 dimensional model of \cite{Horowitz:2008bn}. 
Since the conductivity is directly related to the density of energy eigenstates of the charge carriers,
the presence of peaks gives interesting information about the energy levels of the condensed matter system.
As $T\to 0$, these peaks become very narrow and more pronounced.
We have explained (following the idea of \cite{Horowitz:2009ij}) how the number of peaks increases with the height of the effective potential in a related Schr\"odinger problem. The examples considered in section 5 suggest  that there might be models
in the family parametrized by the couplings $G(\eta)$, $U(\eta)$ and $J(\eta)$ where the potential barrier tends to infinity as $T\to 0$, thus leading to a hard gap and an infinite number of peaks.
This would lead to the intriguing possibility that the dual condensed matter system
would  have a
density of states given by  an infinite sum of delta functions at different frequencies. Such behavior is more typical of systems with an external magnetic field, which breaks the Fermi surface 
leading to a discrete structure for the density of energy eigenstates given by the Landau levels. 
Since in the present case there is no magnetic field, in such scenario the discretization of the spectrum might be due to some strong coupling effect.
Another, possibly more likely, scenario is that, with the back-reaction effects incorporated, all models may always have finite number of peaks,  that is, any model with finite  $q$ 
would have a finite $V_{\rm max}$ at $T=0$. Therefore, as $T\to 0$,  the conductivities would assume a form similar to the $T=0.24$ curve in fig. 3a, namely no hard gap but extremely  suppressed at low temperatures,  a few peaks and then the DC  plateau at 
sufficiently high frequencies. Even within this more conservative scenario, the presence of this finite number of sharp resonances and the gap to the DC plateau  becoming wider at lower temperatures
 is quite striking and begs for a condensed matter explanation.
Investigating this problem in detail requires going to very small temperatures where our numerical approximation with no back-reaction is not reliable.

\medskip

In the second part of the paper, we investigated the  Hall effect induced by
a generalized theta term of the form $\Theta(\eta )\ F\wedge F$. 
Taking couplings with $\Theta (0)=0$, the effect appears only in the  condensed phase.
The direct and Hall conductivities exhibit quite different features.
The direct conductivity has a gap, i.e. it is exponentially small at $\w<\w_g$, and a delta function at $\w=0$
representing the infinite DC conductivity.
The Hall conductivity has no gap and,
at $\w=0$, it has a non-vanishing, finite value. 
We have also studied the temperature dependence of this finite DC Hall conductivity, which goes to zero as $T\to T_c$ with a critical exponent that depends on $\Theta(\eta )$.
This type of Hall effect (also studied in a different context in \cite{Gauntlett:2009bh})) does not require an external magnetic field. Since there are  condensed matter examples where the Hall effect
takes place without any external magnetic field (two examples are the topological Hall effect \cite{Bruno} and the anomalous (spontaneous) AC Hall conductivity \cite{Lutchyn}, the latter being a manifestation of time-reversal symmetry breaking and emerging in a superconducting state),
it would be interesting to see 
if  the generalized ``theta"  interaction might effectively capture the physics of these real systems.

\section{Acknowledgements}

We are grateful to C. Herzog for useful discussions. S. F and D. R-G. would like to thank A. Garc\'ia-Garc\'ia for previous collaboration on related topics. D. R-G. would like to thank the Physics Department at University of Barcelona for warm hospitality while this work was initiated. S. F. is supported by the National Science Foundation under Grant No. PHY05-51164. D. R-G. acknowledges financial support from the European Commission through Marie Curie OIF grant contract
No. MOIF-CT-2006-38381, Spanish Ministry of Science through the resarch grant No. FPA2009-07122 and Spanish Consolider-Ingenio 2010 Programme CPAN (CSD2007-00042).
F.A. is supported by a MEC FPU Grant No.AP2008-04553.
J.R. acknowledges support by MCYT  Research Grant No.  FPA 2007-66665 and Generalitat de
Catalunya under project 2009SGR502.

\end{document}